\documentclass[structabstract]{aa}   
\usepackage{graphicx} 
\usepackage{color} 
\usepackage{natbib}
\bibpunct{(}{)}{;}{a}{}{,}
\def\slantfrac#1#2{\hbox{$\,^#1\!/_#2$}}
\newcommand{\dd}{\, {\rm d}}	    %for an upright d 
\newcommand{\disc}{{\rm disc}}	    %for the disc subscript
\newcommand{\rcyl}{r_{\rm cyl}}	    %for the cylindrical radius 
\newcommand{\teff}{T_{\rm eff}}	    %for the effective temperature
\newcommand{\K}{\, {\rm K}}	    %for a nice Kelvin
\newcommand{\nm}{\, {\rm nm}}	    %for a nice nanometre
\newcommand{\mm}{\, {\rm mm}}	    %for a nice millimetre
\newcommand{\cm}{\, {\rm cm}}	    %for a nice centimetre
\newcommand{\mum}{\ensuremath{\, {\rm \mu m}}}	    %for a nice micrometre
\newcommand{\yr}{\ensuremath{\, {\rm yr}}}	    %for a nice yr
\newcommand{\msun}{\, {\rm M}_\odot}	    %for a nice M sun
\newcommand{\rsun}{\, {\rm R}_\odot}	    %for a nice L sun
\newcommand{\au}{\, {\rm AU}}	    %for a nice AU

\begin{document}    
\title{Observing dust settling and coagulation in circumstellar discs} 
\titlerunning{Dust grain growth and settling in circumstellar discs}
  \subtitle{Selected constraints from high resolution imaging}
    \author{J. Sauter
          \inst{1}
          \and S. Wolf\inst{1}
          }

   \institute{Institut f\"ur Theoretische Physik und Astrophysik, Universit\"at zu Kiel, Leibnizstr. 15, 24118 Kiel, Germany
              \\ \email{jsauter@astrophysik.uni-kiel.de}
  }
 
  \date{Recieved 30 March 2010 / Accepted 18 November 2010}

  \abstract
  % context
  {
    Circumstellar discs are expected to be the nursery of planets.
    Grain growth within such discs is the first step in the planet formation process in the core-accretion gas-capture scenario.
  }
  % aims
  {
    We aim at providing selected criteria on observational quantities derived from multi-wavelength imaging observations that allow to identify  dust grain growth and settling.
  }
  % methods
  {
    We define a wide-ranged parameter space of discs in various states of their evolution.
    Using a parametrised model set-up and radiative transfer techniques we compute multi-wavelength images of discs at different inclinations.
      }
  % results heading
  {
    Using millimetre and sub-millimetre images we are in the position to constrain the process of dust grain growth and sedimentation.
    However, the degeneracy between parameters prohibit the same achievement using near- to mid-infrared images.
    Using face-on observations in the N and Q Band, the sedimentation height can be constrained.
  }
  % conclusions 
  {} 
\maketitle
%
%________________________________________________________________

\section{Introduction}
As the existence of extrasolar planets became evident in the last century the then established understanding of the formation of planetary systems also needed serious revision.
Since then the most accepted evolutionary theory is the so-called ``core-accretion gas-caption'' scenario, featuring a build-up of planet cores first and gas-infall on sufficiently massive cores later. 
In the evolution of T~Tauri stars, circumstellar discs are the natural precursor of a planetary system.
These discs provide both, dust and gas, from which planets are expected to be forming \citep[e.g.][]{2006plfo.book.....K}.
Key issues in the process of planetary core formation are dust grain growth and settling, which start the formation process of planetesimals.
 
Dust grains in the interstellar medium (ISM) can be described by a grain size distribution with a maximum grain size of $a_{\rm max} = 250 \nm$ \citep{1977ApJ...217..425M}.
It has been shown by \cite{2004ApJ...609..826K} that amorphous grains, which account for more than $98\%$ of the dust in the ISM, are smaller than $100\nm$ after contradicting findings by \cite{2001ApJ...550L.201W} and \cite{2003ApJ...594..347D} from X-ray scattering observations.
Molecular clouds are the densest regions of the ISM.
Circumstellar discs form in the stellar genesis by conservation of angular momentum from a parent interstellar molecular cloud that collapses due to instability to its own gravity.
Growth of dust grains starts already in the molecular cloud as observed by e.g. \cite{1980ApJ...235..905W} and modelled by \cite{1993A&A...280..617O} and \cite{1994ApJ...430..713W}.
\cite{2009A&A...502..845O} showed that little coagulation can be expected if the cloud lifetime is limited by the typical free fall timescale in the order of magnitude of $\sim10^5\yr$ as the timescale of coagulation is of the order $\sim10^7\yr$.
In contrast, the dust grain coagulation timescale in discs is $\sim10^4\yr$ \cite[e.g.][]{2005A&A...434..971D}.
 
Evidence for grain growth for T~Tauri stars is numerous and provided for example by spectroscopy at mid-infrared wavelengths \citep{2003A&A...412L..43P,2007ApJ...659..680K}.
Also, low values for the spectral index of the emissivity in the millimetre regime tracing the densest disc regions close to the mid-plane, $\beta$, indicates grain growth \citep{1991ApJ...381..250B}. 

In the evolutionary picture of circumstellar discs, dust grains grow from sub-micrometre sizes to millimetre-sizes via coagulation.
This hit-and-stick process depends on numerous properties of the dust grains as for example their morphology and their relative velocities.
When a dust grain has reached a certain size, its coupling to the gas will weaken and allow it to settle closer to the mid-plane.
Effectively, this leads to a mass-transfer from upper disc regions towards the mid-plane \citep{2006ApJ...638..314D}.
\cite{2004A&A...421.1075D} show that as the disc is depleted in its upper layers, implications to the disc's thermal structure and optical depth profile are inevitable.

According to \cite{1993Icar..106..102C}, dust grains smaller than $\sim 1\,{\rm cm}$ in circumstellar discs can be described within the Epstein regime. 
Small dust grains couple much better to their gas environment than large grains.
Circumstellar discs are understood to be pressure-stabilised against the gravity of the star perpendicular to the disc mid-plane.
As a consequence, larger grains experience a much stronger gravitational force along this direction than smaller grains.
This leads to a settling of large dust grains towards the mid-plane.
An in-depth treatment of the dust layer in a circumstellar disc up to the onset of hydrodynamical stability can be found in \cite{2004ApJ...608.1050G}.
They argue that for a more complete analysis of the situation besides settling, turbulent motion of the gas needs to be considered as well.

Due to the turbulent motion of the gas in deeper disc layers dust grains effectively settle in average at maximum to their respective sedimentation height.
Compared with a disc life time of $\sim10^6 {\rm yr}$ \citep{2009ApJ...701.1188S}, sedimentation is a rather quick process, especially in the upper disc layers.
Dust settling depletes these layers in some $10^3 {\rm yr}$ to $10^4 {\rm yr}$ \citep{2004A&A...421.1075D}.

The physical effect that controls the time scale for the observables to change is grain growth:
As soon as a grain has grown to larger sizes, it will settle almost immediately.

The effects of grain growth on observable quantities of circumstellar discs can be seen by examining the luminosity of a single dust grain and populations thereof.
The luminosity of a single grain with radius $a$, material dependent absorption coefficient $Q_{\rm abs}$ and the black-body luminosity $B(T)$ of a given temperature $T$ at a certain wavelength $\lambda$ can be estimated as
\begin{equation}
  L_\lambda\sim a^2\,Q_{{\rm abs}_\lambda}\, B_\lambda(T).
  \label{eqn_lum}
\end{equation}
The total luminosity of a dust population built from these grains is then given by
\begin{equation}
  L_{{\rm tot}_\lambda} \sim \frac{Q_{{\rm abs}_\lambda}}{a}
\end{equation}
if one uses the specific material density $\rho_{\rm grain}$, the number of grains $n$, and neglects any temperature dependency of $Q_{\rm abs}$.
The scattering properties of dust grains are affected as well \citep{Mie}.
 
In addition to the changing optical properties of the dust grains grain sedimentation results in the modification of the disc density structure.
As a consequence, the spatial structure of the optical depth $\tau_\lambda$ changes, yielding different temperature profiles in the disc and re-emission features of the disc as a function of its evolving state.
The altered optical depth structure also has implications for the morphology of scattered light images.

It is the goal of this study to investigate if the dust grain growth and sedimentation induced change of the disc structure can be observed.
This will provide ideal means to verify current dust-coagulation models as well as allow to constrain the level of turbulence within circumstellar discs.
However, the prediction and interpretation of observable quantities is rather complex.
The optical properties of the dust together with the specific density distribution of the disc result in a highly non-linear behaviour of the respective observables including numerous ambiguities.
To cope with the complexity of the radiative transfer calculations on the model side, Monte Carlo-based codes have proven to be quite successful in the last decade.
On the observation side, the careful interpretation of spectral energy distributions (SED) and images taken at different wavelengths offer a handle on the nature of circumstellar discs.

Generally, the SED of more evolved and settled discs display a smaller excess at longer wavelengths.
This is parallelled in the common classification of classical T~Tauri stars (CTTS) as introduced by \cite{1984ApJ...287..610L}.
However, the Lada classification evaluates fluxes at wavelengths of $2.2\mum$ -- $10\mum$ while the flux decrease in the SED of settled discs occurs from the near-infrared (NIR) to millimetre wavelengths \citep[e.g.]{2005A&A...434..971D}.
\cite{2009A&A...497..379M} showed that the strength and shape of the $10\mum$ feature in T~Tauri stars are related to grain growth.
They also show that disc flaring and grain sizes are slightly related.
\cite{2006ApJ...638..314D} states essential effects of grain growth and dust sedimentation on the SED distribution of the disc can be summarised as follows:
\begin{itemize}
  \item More settled discs feature a smaller far-infrared (FIR)  excess.
  \item More settled discs yield lower midplane temperatures and consequently less flux in the millimetre regime.
  \item Discs seen edge-on lack flux at intermediate wavelengths as a result of high optical depth at short wavelengths.
    A higher degree of sedimentation yields less extinction of the stellar light.
\end{itemize}
 
However, when fitting SEDs to observations, model parameters are often degenerate \citep[e.g.][]{1994A&A...287..493T,1996ApJ...469..366B,2009A&A...502..367S}.
Further, the SED is a spatially integrated observable of a disc.
Hence, it can constrain them only very weakly as demonstrated by \cite{2007ApJS..169..328R}.
 
The effects of dust settling on images of protoplanetary discs in the optical wavelength range are discussed by \cite{2004A&A...421.1075D}. 
\cite{2003ApJ...588L.113M} report the first resolved image of an edge-on disc in the N band and demonstrate that at this wavelength the extent of the scattered light nebula is very sensitive to the grain size distribution.
Experience from modelling individual discs have shown that multi-wavelength, spatially resolving images are needed to allow constraining of the disc density structure and dust grain properties \citep[e.g.][]{2003ApJ...588..373W,2008A&A...489..633P,2009A&A...505.1167S,2010ApJ...712..112D}.
While millimetre observations are sensitive only to radiation being emitted from dust in the densest region within the disc, NIR images are dominated by scattered stellar light from dust in the circumstellar envelope and the upper, optically thin disc layers.
These observations trace different physical processes (scattering/re-emission) in different regions of the circumstellar environment, but they are both strongly related to the dust properties in the system.
\cite{2007prpl.conf..523W} discuss the interpretation of images of multi-wavelength observations in detail.
For example, the disc mass and the dust opacity cannot be disentangled as only their product is observable.
Effects of parameters such as the disc dust mass or the inner and outer radii on images have been studied, both modelling single objects and compilations \citep{2008A&A...489..633P,2009A&A...505.1167S,2009ApJ...704..496B,2007ApJ...659.1637S}. 
Stratified disc models with large dust grains close to the midplane and small grains in the upper layers and an envelope have been used to reproduce observations of several discs such as \cite{2010ApJ...712..112D,2009A&A...505.1167S,2008A&A...489..633P,2003ApJ...588..373W}.
In these studies a multi-wavelength approach is used in order to constrain disc layers with different grain sizes.
Distinct objects have been modelled with different dust models.
For instance, some authors include graphite \citep{2003ApJ...588..373W} and some do not \citep{2008A&A...489..633P}.
In the case of CB~26, this choice has implications for the maximum grain size \citep{2009A&A...505.1167S}.
  
However, predictive studies analysing the impact of grain growth and sedimentation on observable quantities in a large wavelength range have only considered the effects on the SED \citep[e.g.][]{2006ApJ...638..314D,2004A&A...421.1075D} so far.

As the next generation of telescopes and interferometers with high spatial resolution will commence in the not so distant future, it is vital to investigate whether and how grain growth and settling can be constrained by spatially highly resolved images of circumstellar discs.

The present paper provides a first approach in addressing the need for a general investigation of the effects of dust grain growth and sedimentation on images from the near-infrared to the millimetre-regime.
We identify indicators that allow the classification of observations and point out directions of interest that are worth a more detailed approach in subsequent studies.

Addressed key questions are:
\begin{enumerate}
  \item What are the general effects of the grain growth induced mass relocation on images of a disc seen at different inclinations and at different observing wavelengths?
  \item Which tracers of dust grain growth and sedimentation are provided by edge-on millimetre images in general?
  \item What can be learned by combining mid-infrared images at various inclinations?
  \item How can grain growth and sedimentation be constrained by near-infrared images?
\end{enumerate}
The present study will necessarily invoke several simplifications.
Modelling of images of an individual object will provide a better insight in its particular evolutionary stage.
Nevertheless, this study aims at providing a first, non-comprehensive guide into the field of imaging dust grain growth and sedimentation.
 
This paper is organised as follows:
 Section \ref{sec_method} introduces the model, the parameters and the methodology employed.
In  Section \ref{sec_results} we illustrate the general effects of dust grain growth and sedimentation in images for a fiducial model.
We also present tracers for grain growth and sedimentation in millimetre and mid-infrared images that are distinctive from other disc parameters.
In Section \ref{sec_discussion} we discuss cases for which this fails.

\section{Method}
\label{sec_method}
To investigate the effects of dust grain growth and sedimentation on observable quantities, four steps are needed:
First, a model of a circumstellar disc has to be set up that includes the effects of dust settling and coagulation.
Secondly, for all model parameters it has to be determined, whether they shall be set to some typical value or if they should be allowed to vary within a certain range.
In the third step, radiative transfer simulations for every model have to be performed.
As a concluding fourth step, the results of the radiative transfer simulations have to be investigated.In this Section, we give a description of the first three steps.

\subsection{The model}
\label{sec_model}
The basis of all our models is a parametrised disc density distribution.
The dust density distribution $\rho_\disc$ of such a disc can be written as
\begin{equation}
   \rho_\disc(\vec{r})= \rho_0 \left( \frac{R_{100}}{\rcyl} \right)^\alpha \exp \left(-\frac{1}{2}\left[ \frac{z}{h}\right]^2 \right)
   \label{eqdiscden}
\end{equation} 
where $z$ is the usual axial coordinate with $z=0$ corresponding to the disc midplane and $\rcyl$ is the radial distance from the $z$--axis.
The quantity $\rho_0$ is determined by the mass of the entire disc, $R_{100} = 100\au$ is the radial normalisation constant, and $h$ the local scale height: 
\begin{equation}
   h(\rcyl) = h_0 \left( \frac{\rcyl}{R_{100}} \right)^\beta.
   \label{eqscaleheight}
\end{equation}
Here, $\alpha$, $\beta$, and $h_0$ in Eqn. (\ref{eqscaleheight}) are geometrical parameters.
This disc model has served well already in describing observed discs, such as HH~30 \citep{2002ApJ...564..887W}, the Butterfly-Star IRAS 04302+2247 \citep{2003ApJ...588..373W}, HK Tau \citep{1998ApJ...502L..65S}, IM Lupi \citep{2008A&A...489..633P}, HV Tau \citep{2003ApJ...589..410S}, and CB~26 \citep{2009A&A...505.1167S}.

To incorporate dust settling and coagulation, this ansatz has to be extended.
A considerable amount of work has been done to understand the physics of dust grain coagulation and settling, including extensive numerical simulations on the topic, as well as dust coagulation studies in laboratories. 
A comprehensive treatment can be found in a series of papers by
\cite{2006ApJ...652.1768B,2008ApJ...675..764L,2009ApJ...696.2036W,2009ApJ...701..130G}.
Detailed models of dust grain growth and settling and the resulting vertical disc structure can be found in \cite{2004A&A...421.1075D}.
An analysis of the radial drift of dust grains can be found in \cite{2008A&A...480..859B} and \cite{2010arXiv1002.0335B}. 
However, in the present work, we do not include the detailed picture of the dust growth process developed there.
Instead, we use the parametrised ansatz that has been developed and successfully employed by \cite{2006ApJ...638..314D}.This enables us to formulate a parameter space that includes not only basic parameters of the coagulation/sedimentation process but also important observational parameters that do not influence the disc physics but the way observations perceive them.

\begin{figure}
  \resizebox{\hsize}{!}{
    \includegraphics[width=\textwidth]{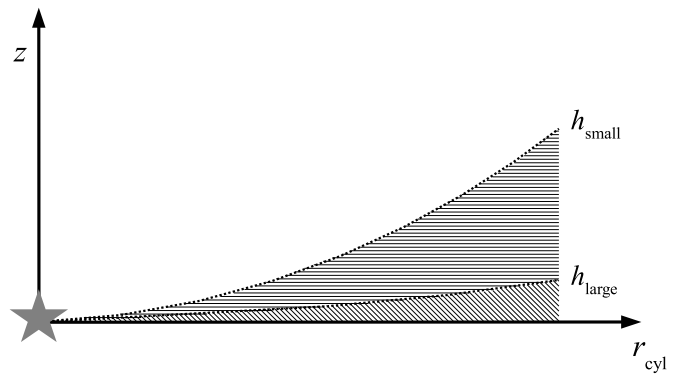}
   } 
  \caption{
    Sketch of the general model setup.
    Shown are the scale heights $h_{\rm small}$ and $h_{\rm large}$ of the two dust populations 
    with $\tilde{h}=\frac{h_{\rm small}}{h_{\rm large}}$.
    Note that the scale height is not the upper border of a dust population.
    }
  \label{sketch}
\end{figure}
To describe grain growth and settling, our model features two distinct dust populations which differ only in the maximum grain size.
For each population, the minimum grain size is the same as found in the ISM: $a_{\rm min}=5\,{\rm nm}$.
The first population has the same maximum grain size as commonly assumed for the interstellar medium: $a_{\rm max}=250\,{\rm nm}$ \citep{1977ApJ...217..425M}.
This dust population is considered the reservoir of small dust grains from which the second population of larger grains is fed.
The second component features a maximum grain size of $1\,{\rm mm}$.

As described by \cite{2004A&A...421.1075D}, the height to which a dust grain settles does not only depend on its size but also on the level of turbulence present in the disc.
To model dust settling both grain populations feature different scale heights that are related via
\begin{equation}
  \tilde{h}=\frac{h_{0,{\rm small\ grains}}}{h_{0,{\rm large\ grains}}}.
  \label{eqntildeh}
\end{equation} 
Here, $h_{0,{\rm small grains}}$ and $h_{0,{\rm large grains}}$ are the respective scale heights.
Figure~\ref{sketch} depicts the model concept.
 
In the initial state both populations have the same dust-to-gas ratio.
However, once the disc evolves towards the grown and settled case, the small dust grain population gets depleted and mass is transferred to the large dust grain population.
The model of \cite{2006ApJ...638..314D} parametrises the relative weighting of the two populations by the depletion parameter $\epsilon$ which is the relation of the dust-to-gas ratio in the small grain population $\zeta_{\rm small}$ to the default dust-to-gas ratio $\zeta=\frac{M_{\rm Gas}}{M_{\rm Dust}} = 100$:
\begin{equation}
  \epsilon = \frac{\zeta_{\rm small}}{\zeta}. $$
\end{equation}

As no accretion by the central star nor mass loss to the ambient space is considered, the dust-to-gas-ratio $\zeta_{\rm large}$ of the large grain population has to be altered in order to keep $\zeta$ at its initial value.
The values of $\zeta_{\rm large}$ as a function of $\zeta_{\rm small}$ are taken from \cite{2006ApJ...638..314D}.
Also, we keep $\epsilon$ independent of the radial position in the disc.

\subsection{The parameter space}
Our model features among intrinsic parameters the observing wavelength and the disc inclination as external parameters.
We now discuss their value and range. 
 
\subsubsection{Stellar heating}
The stellar radiation heats the dust which then itself re-emits at longer wavelengths.
We neglect accretion and turbulent processes within the disc as another possible intrinsic energy source.
In this sense the disc in our model is passive.
We employ an ``average''  TTauri star as given by the survey of \cite{1998ApJ...492..323G}. 
This star has a radius of $r=2\rsun$ and an effective surface temperature of $\teff=4000 \K$.
Assuming the star to be a blackbody radiator, this yields a luminosity of  $L_* = 0.92 {\rm L}_\odot$.

\subsubsection{Dust properties}
We assume that the gas within the disc is optically thin and limit ourselves to radiative transfer through the dust and neglect line emission and absorption by the gas.
This is justified as the dust is the dominant coolant of the disc.
Hence, thermal equilibrium obtained in radiative transfer calculations based on dust only will provide a reliable description of the disc's thermal structure.
This assumption is based on the work of \cite{2004ApJ...615..991K}, who showed, that dust and gas temperature in a circumstellar disk around a T~Tauri star are well coupled in the regions which are sufficiently dense to be traced with current and near-future (sub)millimetre observatories.

We assume the dust grains to be homogeneous spheres.
Real dust grains are expected to feature a much more complex and fractal structure.
As discussed by \cite{2002ocd..conf....1V}, chemical composition, size and shape of dust grains cannot be determined separately, but only as a combination.

We keep the dust composition of silicate graphite material fixed in order to avoid degeneracies.
For the optical data we use the complex refractive indices of ``smoothed astronomical silicate'' and graphite as published by \cite{2001ApJ...548..296W}. 
For graphite we adopt the common ``$\frac{1}{3} \ - \ \frac{2}{3}$'' approximation.
As has been shown by \cite{1993ApJ...414..632D}, this graphite model is sufficient for extinction curve modelling.
Applying an abundance ratio from silicate to graphite  of $1 \times 10^{-27} \cm^3 {\rm H}^{-1}$ : $1.69 \times 10^{-27} \cm^3 {\rm H}^{-1}$, we get relative abundances of $62.5 \%$ for astronomical silicate and $37.5\%$ graphite ($\frac{1}{3} \epsilon_\parallel$ and $\frac{2}{3}\epsilon_\perp$).
As the grain size distribution we assume a power law of the form
\begin{equation}
  n(a) \dd a \sim a^{-3.5}  \dd a \quad {\rm with} \quad  a_{\rm min} < a < a_{\rm max}.
  \label{eqgraindist}
\end{equation} 
Here, $a$ is the dust grain radius and $n(a)$ the number of dust grains with a specific radius.
For each dust grain ensemble, 1000 logarithmically equidistantly distributed grain sizes within the interval $[a_{\rm min} : a_{\rm max} ]$ have been taken into account.
For simplicity reasons, we assume Eqn. \ref{eqgraindist} to be valid in the whole disc for each grain population individually.
\cite{1994ApJ...422..164K} argued that in order to achieve good fit to U, B, and V linear polarisation measurements, the grain size distribution has to depart from the power law.
However, the grain size distribution in a specific region of the circumstellar disc depends on the grain evolution there which depends on a variety of factors.
First, grains are understood to grow fast in the denser regions of the disc.
Secondly, mixing and fragmentation processes will also influence the grain size distribution.
With the model employed in this paper utilising two different grain populations we consider effects of the vertical dust grain settling.
The \emph{overall} grain size distribution then deviates from a simple power law accordingly.
Finally, we assume an average grain mass density of $\rho_{\rm grain} = 2.5 \,{\rm g}\,{\rm cm}^{-3}$.
  
\subsubsection{Parametrisation of growth and settling: $\epsilon$ and $\tilde{h}$}
In the framework of our model we follow the parametric values given by \cite{2006ApJ...638..314D} for the depletion parameter $\epsilon$ and the relative sedimentation height $\tilde{h}$.
For coagulation and settling induced deviations from the default mass-to-gas relation of the small grain population we have $\epsilon = 1.0$, $0.1$, $0.01$, and $0.001$.
The quantity $\tilde{h}$ describes the level of turbulence in the disc.
As this level is generally not known, we adopt three values for $\tilde{h}_0$, scattered around the \cite{2006ApJ...638..314D} value of 10.
Namely we have  $\tilde{h}=8$, 12, and 20.

\subsubsection{Disc flaring and surface density}
In our model (Eqn. \ref{eqdiscden}) two exponents $\alpha$ and $\beta$ control the density distribution of the disc.
The exponent $p$ of the surface density $\Sigma=\Sigma_0r^p$ is directly related to $\alpha$ and $\beta$ via $p=\alpha -\beta$.
Most observed circumstellar discs can be well described by a relatively small range of values for $\alpha$ and $\beta$.
Good agreement between model and observations are found for values of $\beta = \frac{5}{4}$ and for $\alpha$ by the relation $\alpha = 3\,\left(\beta-\frac{1}{2}\right)$, which is a result of viscous accretion theory \citep{1973A&A....24..337S}.
Thus, these two parameters are also kept fixed in order to reduce the free parameters of the model.
We employ the values as obtained for the Butterfly star \citep{2003ApJ...588..373W}: $\alpha=2.37$ and $\beta=1.29$.

\subsubsection{Inner and outer radial extent of the circumstellar disc}

Earlier and recent modelling efforts have shown a variety of size scales of the inner disc radius.
Some discs are best described using an inner radius as small as the sublimation radius for silicate at $T=1500\,{\rm K}$.
This temperature is reached at a distance of about $0.07\au$ from the T~Tauri star in our model.
An example for this configuration can be found in \cite{2003ApJ...588..373W}.
Other discs, such as HH~30 \citep{2008A&A...478L..31G}, apparently feature a large inner void with radii up to $\sim 45\au$ from the star. 
To incorporate both types of discs, we allow the inner radius to assume the values $r_{\rm inner} =0.1\au$, $1.0\au$, and $10\au$.
The outer radius of the dust disc is also subject to variation.
For discs around T~Tauri stars, disc radii vary from $25\au$ \citep{2009A&A...496..777N} up to $400 \au$ \citep{2000ApJ...545.1034S}.
Hence, we allow in our study the outer disc radius to assume the values $r_{\rm outer}=100\au$, $300\au$, and $400\au$.

\subsubsection{The total disc dust mass}
The values for the total dust mass of the disc span a range of three orders of magnitude:
$m_{\rm dust}=3.2\times 10^{-6}\msun$, $3.2\times 10^{-5}\msun$, and $3.2\times 10^{-4}\msun$.

\subsubsection{Choices of inclination}
The inclination under which a circumstellar disc is observed is a non-intrinsic quantity of the model.
In the case of face-on discs ($\theta=0^\circ$) one has direct observational access to the central star while in the edge-on case the star is typically completely obscured by the dust disc.
On the other hand, the edge-on case might allow to examine the vertical structure better if upper disc layers are sufficiently transparent.
It is one goal of this study, to determine to which extent the vertical disc structure can be determined by observations.
We include both extrema in our study. 

\subsubsection{Observing wavelengths}
To predict observational quantities in a multi-wavelength set-up, we compute images in the near-infrared, in the mid-infrared, and in the sub-millimetre and millimetre regime.
Altogether, this gives us images in the I, J, H, K, L, M, N, and Q Band (central wavelength $\lambda =1.0\mum, 1.25\mum, 1.6\mum, 2.2\mum, 3.4\mum,$ $4.7\mum$, $\lambda=10\mum$, and $ 20\,{\rm \mu m}$) as well as $15\mum$, $ 50\,{\rm \mu m}$, $ 100\,{\rm \mu m}$, $ 350\,{\rm \mu m}$, $ 670\,{\rm \mu m}$, $ 840\,{\rm \mu m}$, $ 1100\,{\rm \mu m}$, and $ 1300\,{\rm \mu m}$.
For computations of observed fluxes a distance of 140\,pc is assumed, representative for the well-studied low-mass star-forming region in the Taurus constellation.

\subsection{Radiative transfer simulations}
For the continuum radiative transfer simulations we make use of the program \texttt{MC3D}\footnote{Available upon request: \texttt{wolf@astrophysik.uni-kiel.de}}
\citep{1999A&A...349..839W,2003CoPhC.150...99W}.
It is based on the Monte-Carlo method and solves the continuum radiative transfer problem self-consistently.
The program uses the temperature correction technique as described by \cite{2001ApJ...554..615B}, the absorption concept as introduced by \cite{1999A&A...344..282L} and the enforced scattering scheme as proposed by \cite{Cashwell}.
Multiple and anisotropic scattering is considered.
The radiative transfer is simulated at 100 wavelengths, logarithmically distributed in the wavelength range  $[\lambda_{\rm min},\lambda_{\rm max}] = [50 \nm, 2.0 \mm ]$.

\subsection{Beyond the model}
Our model setup is based on several simplifications.
The shape of the disc, governed by $\alpha$ and $\beta$ in Eqs. \ref{eqdiscden} and \ref{eqscaleheight} is kept fixed.
Flaring of the disc surface given by $\beta$ has a significant influence on the amount of intercepted stellar radiation, consequently the thermal structure of the disc, as well as the spatial distribution of scattered light.
As discussed by \cite{2007prpl.conf..523W}, this two parameters have also an impact on the width of the dark lane similar as depletion parameter $\epsilon$ (see Sec. \ref{sec_fidu}).
Consequently, the freedom to vary $\alpha$ and $\beta$ harbours a degeneracy in this respect.
Yet, this is not necessarily a draw back: In order to understand the influence of all possible physical processes, careful investigation of each is needed separately.
To do this for the effects of sedimentation on images in the described framework is the goal of this paper.
 
\section{Results} 
\label{sec_results}
We investigate the parameter space of our model for effects on various observational quantities that allow one to constrain dust settling and coagulation.
In particular, we examine the implications of the depletion parameter $\epsilon$ and the scale height relation $\tilde{h}$ in Section \ref{sec_fidu}.
The general result is that dust settling and coagulation as described by the depletion parameter $\epsilon$ can be constrained by various observations if the disc is seen edge-on.
The respective findings are presented in Section \ref{sec_millimeter} and \ref{sec_NQ_faceon}.
Possibilities to constrain the relative scale height $\tilde{h}$ (i. e. the amount of turbulence in the disc), is discussed in Section \ref{sec_NQ_gap}.
These findings hold true only in the parameter space outlined in Section 2, especially the exact value of modelled fluxes or the morphology of images, particularly in the mid-infrared.
Nevertheless, the qualitative behaviour of tracers for grain growth and sedimentation outlined in this Section can be also expected for models with, for instance, different choices for $\alpha$, $\beta$ or $h_0$ in the density distribution (cf. Eqn. \ref{eqdiscden}).

\subsection{The fiducial model}
\label{sec_fidu}
To discuss the effects of dust grain growth and coagulation on observable quantities and as a reference for further investigations we introduce a fiducial model.
Its parameters are: $r_{\rm inner} = 0.1\au$, $r_{\rm outer}=300\au$, $m_{\rm dust}=3.2\times10^{-4}\msun$, $\tilde{h}=8$, and $\epsilon=1.0$.
This approach allows us to evaluate the impact of variations of the individual model parameters on the observable quantities.
Resulting degeneracies are discussed.

\paragraph{Effects of the depletion parameter $\epsilon$}
The depletion parameter $\epsilon$ governs how much mass is transferred from the small to the large dust grain population. 
Therefore, $\epsilon$ can be understood as a parameter for dust grain coagulation and the resulting sedimentation.

The near-infrared images of edge-on discs exhibit a pronounced chromaticity.
In this wavelength regime the disc's midplane is optically thick and shields the central star from direct observation.
The height above the midplane at which the disc becomes transparent to the scattered stellar light constrains the disc's scale height.
 
\begin{figure} 
  \resizebox{\hsize}{!}{   
    \includegraphics[width=\textwidth]{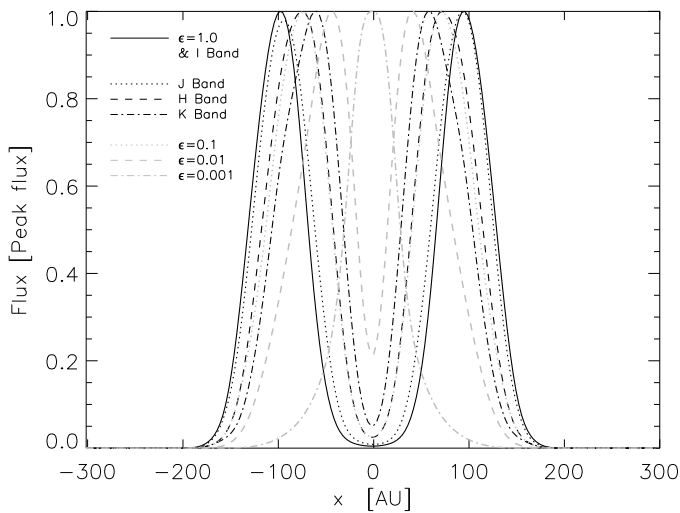}
  }
  \caption{
    Dependency of the width of the dark lane in scattered light images on the depletion
    parameter $\epsilon$ (grey) and the observing wavelength $\lambda$ (black).
    Shown are cuts through images along the disc midplane centred on the projected position of the embedded star.  
   } 
  \label{darklanes}
\end{figure}

Besides the optical properties of the dust particles, the opacity of the disc depends on the amount of dust from both dust populations present in the respective disc layers.
Considering dust settling, the composition of those layers is a function of the degree of depletion.
Thus, the width of the dark lane is not a function of the disc gas scale height alone but must also be a function the depletion parameter $\epsilon$.
The fiducial model exhibits this behaviour.
Fig. \ref{darklanes} shows both, the dependency of the dark lane on the observing wavelength $\lambda$ as well as on the depletion parameter $\epsilon$.
Although the values for $\lambda$ and $\epsilon$ in the plot do not exactly mirror each other
the width of the dark lane can be reproduced adjusting both parameters.

\begin{figure}
  \resizebox{\hsize}{!}{
    \includegraphics[width=\textwidth]{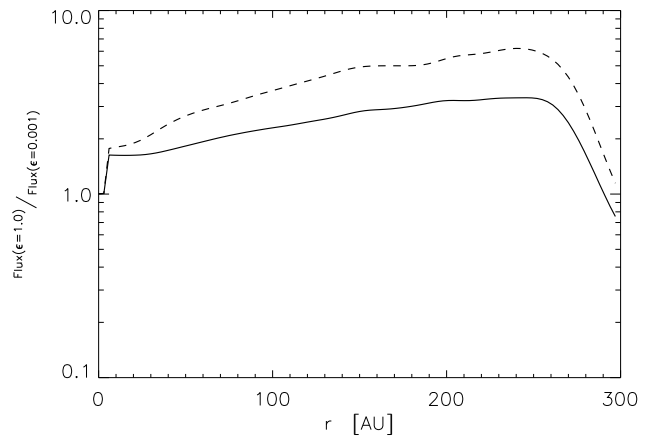}
   } 
  \caption{ 
  Selected flux ratios of the radial brightness profile of the fiducial model seen in the H band face-on.
  The solid line shows the flux ratio between models with $\epsilon=1.0$ and $\epsilon=0.001$.
  The dashed line shows the ratio between fiducial models with $m_{\rm dust} = 3.2\times10^{-4}\msun$ and  $m_{\rm dust} = 3.2\times10^{-6}\msun$.
  }
  \label{plot_faceon_rp_relation}
\end{figure}
\begin{figure}
  \resizebox{\hsize}{!}{
    \includegraphics*[width=\textwidth]{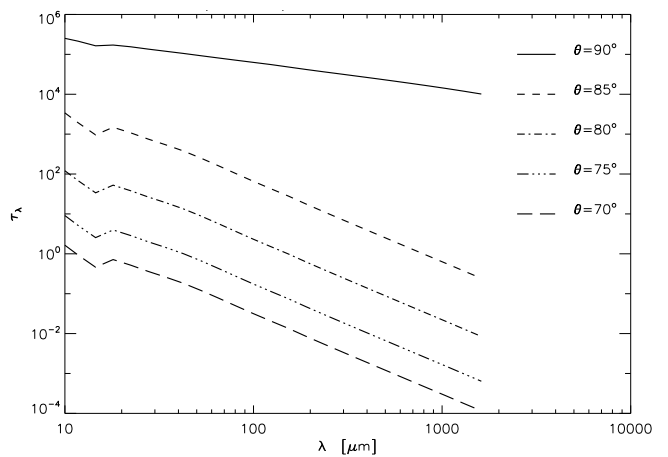}
   } 
  \caption{  
  	Optical depth of the fiducial model along radial paths for selected inclinations as a function of wavelength.
	A high optical depth in the millimetre regime is obtained only for high inclinations while for shorter wavelengths the model is opaque in a wider range.
  } 
  \label{plot_radtau}
\end{figure}
\begin{figure}
  \resizebox{\hsize}{!}{
    \includegraphics*[width=\textwidth]{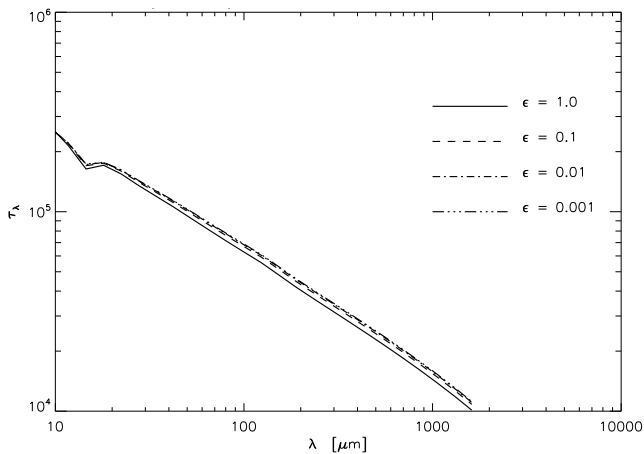}
   } 
  \caption{ 
  	Optical depth of the fiducial model along a radial path at $90^\circ$ inclination as a function of wavelength.
	Only a weak dependency on the depletion parameter $\epsilon$ exists.
  }
  \label{plot_epsilon_radtau_90}
\end{figure}
\begin{figure}
  \resizebox{\hsize}{!}{
    \includegraphics*[width=\textwidth]{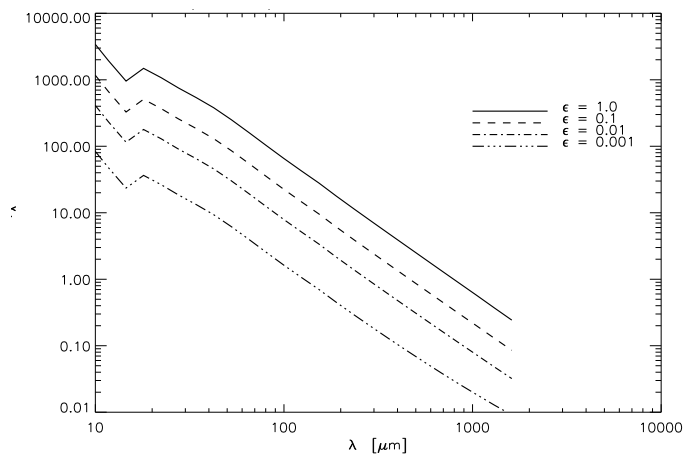}
   } 
  \caption{
  	Optical depth of the fiducial model along a radial path at $85^\circ$ inclination as a function of wavelength.
	Unlike the situation in the exact edge-on case (Fig.~\ref{plot_epsilon_radtau_90}), the optical depth becomes sensitive to the depletion parameter $\epsilon$ by typically half an order of magnitude.
	} 
  \label{plot_epsilon_radtau_85}
\end{figure} 
Fig.~\ref{plot_faceon_rp_relation} shows the relation of the radial brightness profile of the fiducial model with different total dust masses and different depletion parameters $\epsilon$ observed face-on.
For $r=0$, the ratio between the fluxes of the fiducial model with $\epsilon=1.0$ and with $\epsilon=0.001$ 1 since the same stellar properties are used for all models.
For $r>0$, the ratio between the brightness profiles is $\approx 2$ for different depletion parameters $\epsilon$ while it is slightly sloped for different total dust masses.
A variation of the density structure (i.e. $\alpha$ and $\beta$ in Eqn. \ref{eqdiscden}) can easily counterbalance the effects shown in the plot.
For inclinations $\theta=60^\circ$ and $\theta=0^\circ$ the dependence of the brightness distribution on $\epsilon$ is almost lost.

The first reason for this ambiguity is that as the disc is tilted to smaller inclinations, the dark lane vanishes.
This is because the optical depth along the line of sight decreases and more scattered stellar light reaches the observer.

Secondly, as the disc itself is still optically thick in the near infrared, it is still only the upper disc layers that are penetrated by stellar light.
Hence, it is the same physical environment as in the edge-on case that determines the characteristics of the scattered light. 
 
Fig.~\ref{plot_radtau} shows the dependency of the optical depth as a function of the wavelength for selected radial paths.
A high optical depth ($\tau_\lambda \gg 1$) in the millimetre regime is obtained only at high inclinations.
For shorter wavelengths, especially in the near-infrared, the fiducial model is opaque even for $\theta \sim 70^\circ$.
Figures \ref{plot_epsilon_radtau_90} and \ref{plot_epsilon_radtau_85} show the dependency of $\tau_\lambda$ along a radial path on the depletion parameter $\epsilon$. 
For the edge-on orientation $\theta = 90^\circ$ the dependency $\tau_\lambda(\epsilon)$ is very weak (Fig. \ref{plot_epsilon_radtau_90}).
Yet, $5^\circ$ away from the edge-on orientation, the optical depth is much more sensitive to the depletion parameter $\epsilon$.
  
For images taken in the sub-millimetre and millimetre regime no distinctive effect is observed.
While absolute flux values change as the large dust grain population becomes more massive, no relative 
effect is seen either with respect to the location within one image or as a function of wavelength.

These findings hold true for any inclination we considered in our framework.
Since for smaller values of $\epsilon$ more large grains are present in the system and large grains re-emit more efficiently at larger wavelengths than smaller grains do, one might expect observable effects in the millimetre regime.
For the case of an edge-on disc the results are discussed in Section \ref{sec_millimeter}.
However, if the disc is seen face-on, there are no observable effects. 

\paragraph{Effects of the height relation $\tilde{h}$}
The height relation $\tilde{h}$ as defined in Eqn. \ref{eqntildeh} describes the level of turbulence in the disc as well as the magnitude of the viscosity parameter $\alpha$.
\begin{figure}
  \resizebox{\hsize}{!}{
    \includegraphics[width=\textwidth]{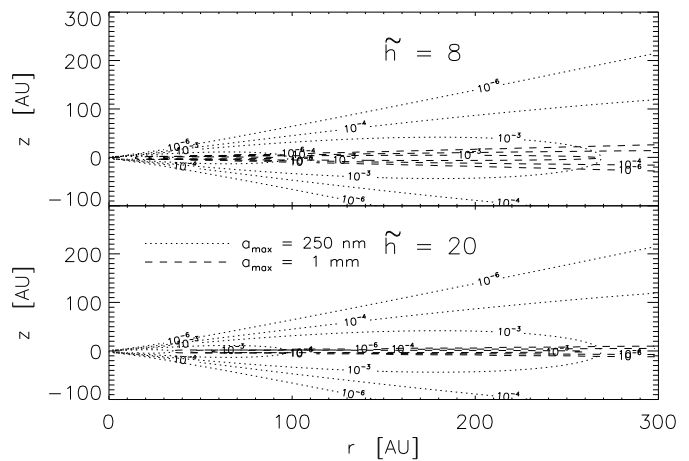}
  }
  \caption{
     Contour lines of the density distribution normalised to the respective peak density of the two grain populations.
     The two panels show the effect of small (\emph{top}) and large (\emph{bottom}) relative scale height $\tilde{h}$.
  } 
  \label{plot_dens}
\end{figure}
Fig.~\ref{plot_dens} illustrates the effects of $\tilde{h}$ on the density distribution.
The density distribution of the small grain population does not depend on $\tilde{h}$.
According to the model set-up the large grain population becomes concentrated in a much smaller space for larger $\tilde{h}$.

The effects of $\tilde{h}$ on images at different inclinations is essentially that of an amplifier of effects
caused by  the depletion parameter $\epsilon$. 
In the case of near-infrared images of edge-on discs, the width of the dark, high-opacity lane is affected.
Larger values for the height relation $\tilde{h}$ result in a dark lane with larger contrast and vice versa.
This can be readily understood in terms of optical depth.
Higher values for $\tilde{h}$ essentially squeezes the large dust grain population into a smaller volume and thus increases the optical depth.
The amplifying effect is stronger for smaller values of $\epsilon$: more mass is packed into the large dust grain population yielding a height optical depth.
The magnitude of the effects caused by the variation of $\tilde{h}$ is considerably small.

\begin{figure}
  \resizebox{\hsize}{!}{
    \includegraphics*[width=\textwidth]{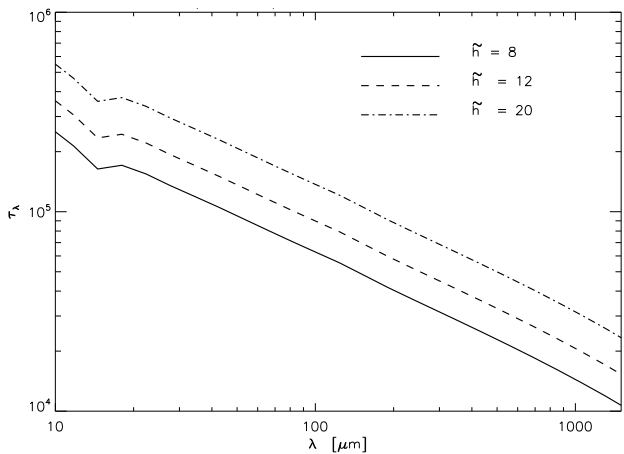}
   } 
  \caption{
  	Optical depth of the fiducial model along a radial path at $90^\circ$ inclination as a function of wavelength.
	Plotted is the dependence on the relative scale height $\tilde{h}$.
	A comparison with Fig.~\ref{plot_epsilon_radtau_85} shows, that the dependence on $\tilde{h}$ is not as pronounced as the dependency on $\epsilon$.
  }
  \label{plot_h-radtau}
\end{figure}
In Fig.~\ref{plot_h-radtau} the optical depth of the fiducial model along a radial path in the edge-on orientation as a function of wavelength is shown.
The comparison with Figs.~\ref{plot_epsilon_radtau_90} and \ref{plot_epsilon_radtau_85} shows that for $\tilde{h}$ the dependency is not as strong as for $\epsilon$.
 
\subsection{Millimetre Images}
\label{sec_millimeter}
Millimetre images are well suited to classify dust settling and coagulation because at these wavelengths the dominant fraction of the radiation has its origin in the thermal re-emission of the dust grains.
Further, the redistribution of the mass also alters the optical depth structure.

\begin{figure}
  \resizebox{\hsize}{!}{
    \includegraphics[width=\textwidth]{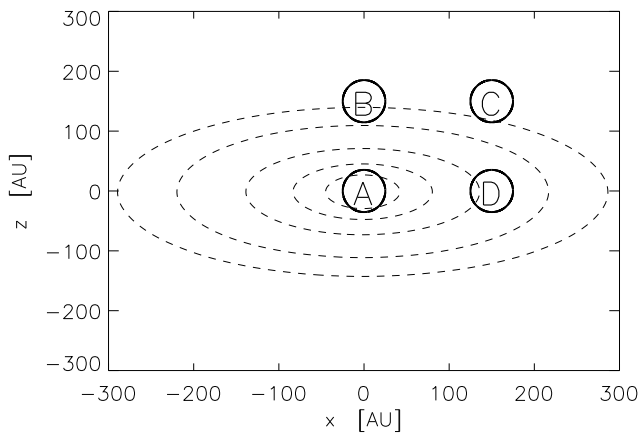}
   } 
  \caption{
    Location of the four regions used to compare intensities.
    The dashed lines are logarithmic contour lines of the brightness distribution of the fiducial model observed at $\lambda=1.3\mm$ and $\theta=60^\circ$.
    }
  \label{fourregions}
\end{figure}
The most illustrative way to see how millimetre images provide means to constrain disc evolution (the changing radial and vertical opacity and thus the reemission structure) is to compare different regions of a disc image,
especially if the disc is seen edge-on.
The regions of our choice are depicted in Fig. \ref{fourregions} and  labelled (cw, starting in the image/disc centre) A, B, C, and D.
One expects that as the disc evolves, the layers distant from the midplane loose more and more mass due to the dust grain growth and settling towards the midplane, the amount of thermal re-emission from those layers decreases as well.
Radiation originating close to the midplane is expected to raise or to stay at the same level, since the growing midplane mass not only yields more emitting grains but also more absorbing grains (important in the submillimeter regime).
Densities values are already orders of magnitude higher close to the disc midplane, the down-raining grains therefore cannot significantly increase the optical depth and thus level any gain in emitter numbers.

\begin{figure}
  \resizebox{\hsize}{!}{
    \includegraphics*[width=\textwidth]{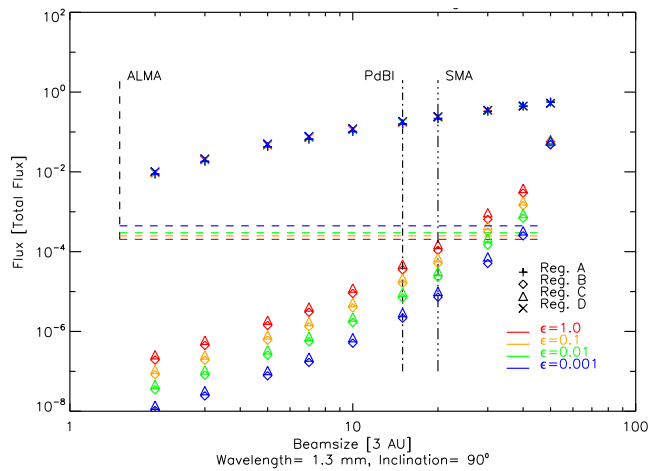}
   } 
  \caption{
    Analysis of image brightness distribution of a circumstellar disk at different evolutionary stages.
    The results for different evolutionary states are coded in colour.
    The shape of any mark denotes the region from whence it is taken.
    Vertical lines indicate resolution limits of selected interferometer.
    Horizontal lines give the ALMA sensitivity limit, the other two instruments do not have the required sensitivity.
    The flux is given in units of the total flux.
    Since this flux depends on $\epsilon$ as well, the sensitivity limit is affected.
    The dependency of the flux in regions B and C on $\epsilon$ is shown.
  }
  \label{result90deg}
\end{figure} 
Fig.~\ref{result90deg} shows exemplarily our findings by relating the flux from the four different regions to the total flux. 
Starting from red for the early, $\epsilon=1.0$ case and ending at blue with a depletion parameter $\epsilon=0.001$.
The shape of any mark denotes the region from whence it is taken.
Plotted is the relative flux from a certain region versus the size of that region.
The latter can be understood as the degree of resolution of the observing instruments.
As an indicator, the maximum resolution  1.3\,mm of various instruments (ALMA, PdBI, SMA) is given as dashed black lines.
Although all three interferometers shown in the plot provide the required resolution to detect the effect, only ALMA provides the required sensitivity.
As the flux is given in units of the total flux of the image, which depends on $\epsilon$, the sensitivity is also a function of $\epsilon$. 
As expected, fluxes taken in the midplane do not vary considerably with the evolutionary state: all four colours are almost atop of each other.
However, for fluxes taken from above the disc midplane, a clear decrease in flux is visible.
From the initial $\epsilon=1.0$ to the final $\epsilon=0.001$, it spans almost two orders of magnitude.
This effect of flux decrease though is limited to the edge-on case, only.

At  inclinations of $\theta=60^\circ, 0^\circ$, the lines of sight are no longer within the midplane or any parallel plane but intersect higher disc layers as well as the disc midplane.
As a result the radiation comes from various  disc layers, representing different stages of dust coagulation and sedimentation.
The amount of flux collected in the four different regions is not a function of the depletion anymore but just of the disc's geometrical thickness.
As the inclination can be determined from image-morphology, these findings are a valuable tracer for dust sedimentation in the edge-on case.
To estimate the robustness of this result, we consider the influence of disc parameters other than the depletion factor $\epsilon$.
We find that although the total flux value from different regions changes, the effect of separation of the flux values in the upper and lower layers of the image with later evolutionary states does not depend on the chosen parameter value.
It is also independent of the observing wavelength in the (sub) millimetre regime. 

\subsection{The disc in N and Q band edge-on}
\label{sec_NQ_faceon} 
We now turn our attention towards the potential of mid-infrared observations in the N and Q bands.
\begin{figure}
  \resizebox{\hsize}{!}{
    \includegraphics[width=0.5\textwidth]{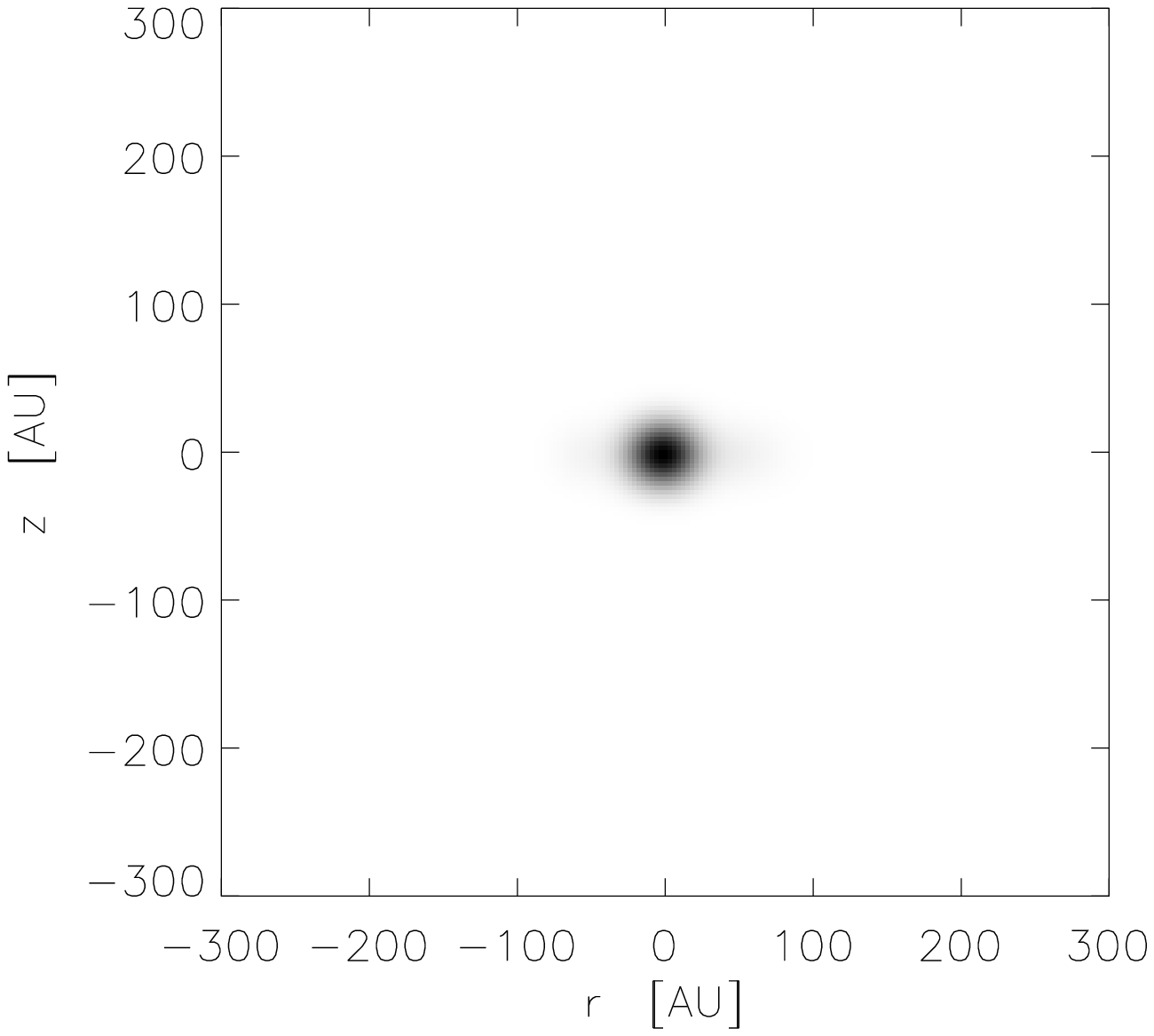}
    \includegraphics[width=0.5\textwidth]{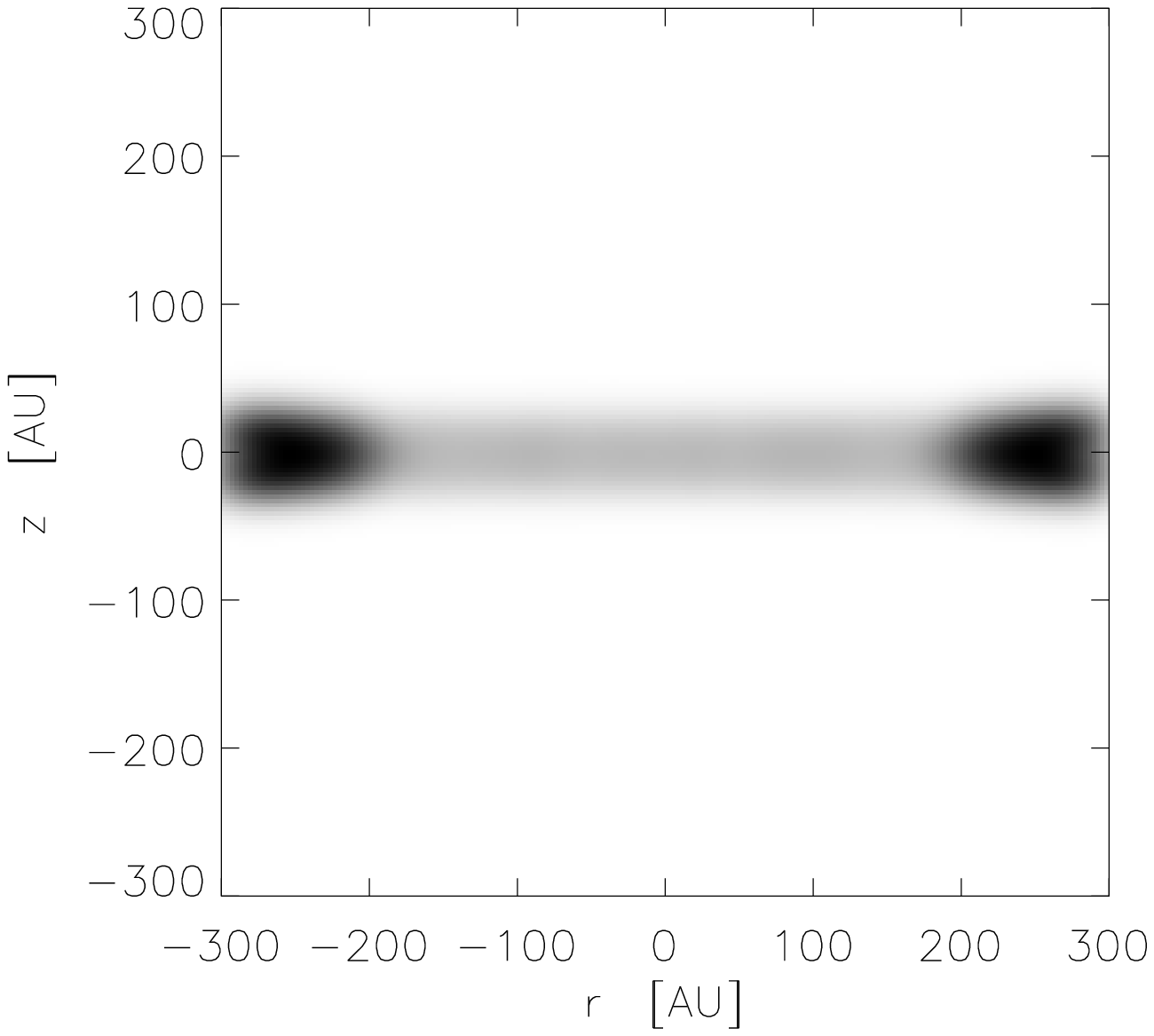}
   } 
  \resizebox{\hsize}{!}{
    \includegraphics[width=0.5\textwidth]{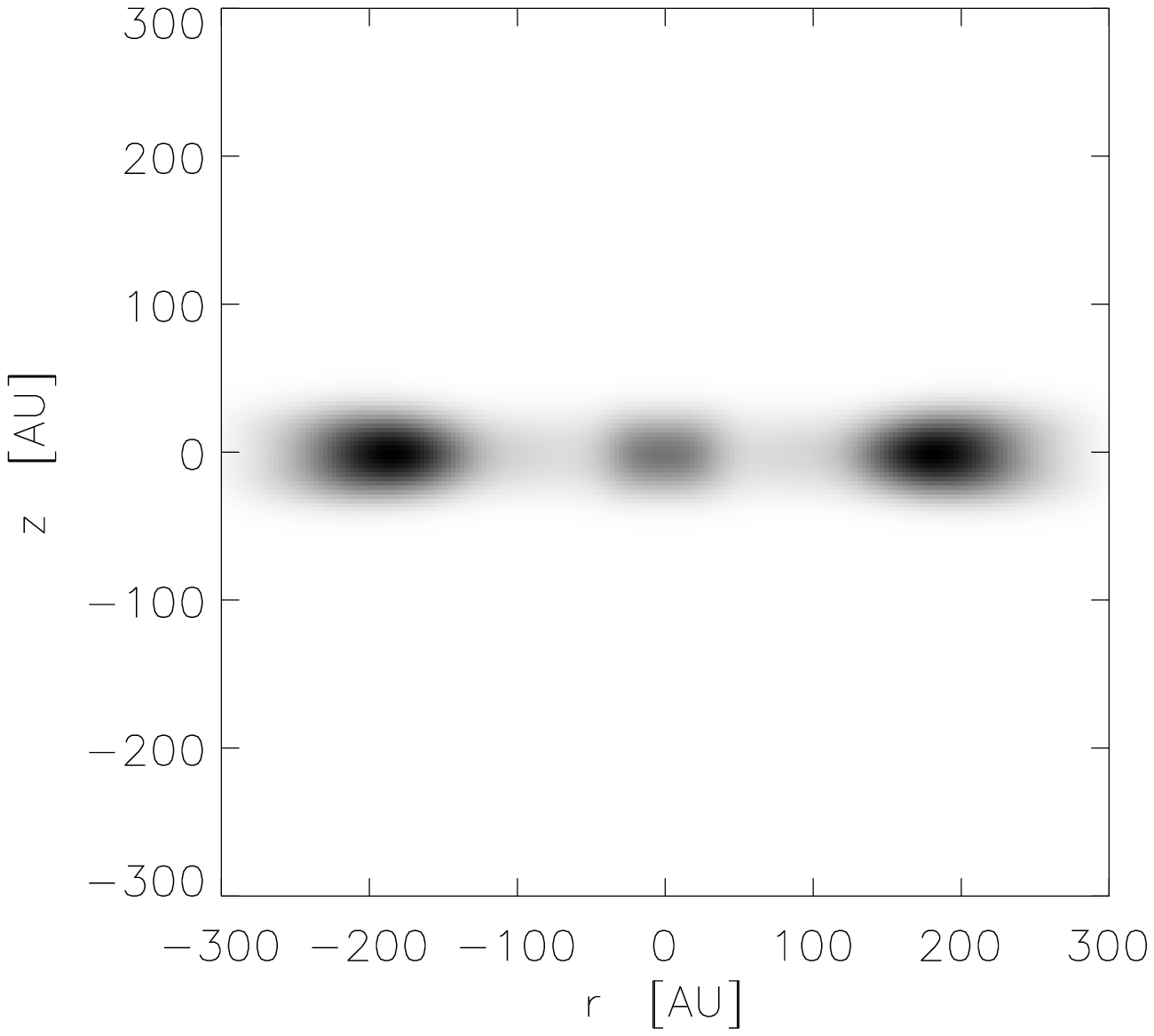}
    \includegraphics[width=0.5\textwidth]{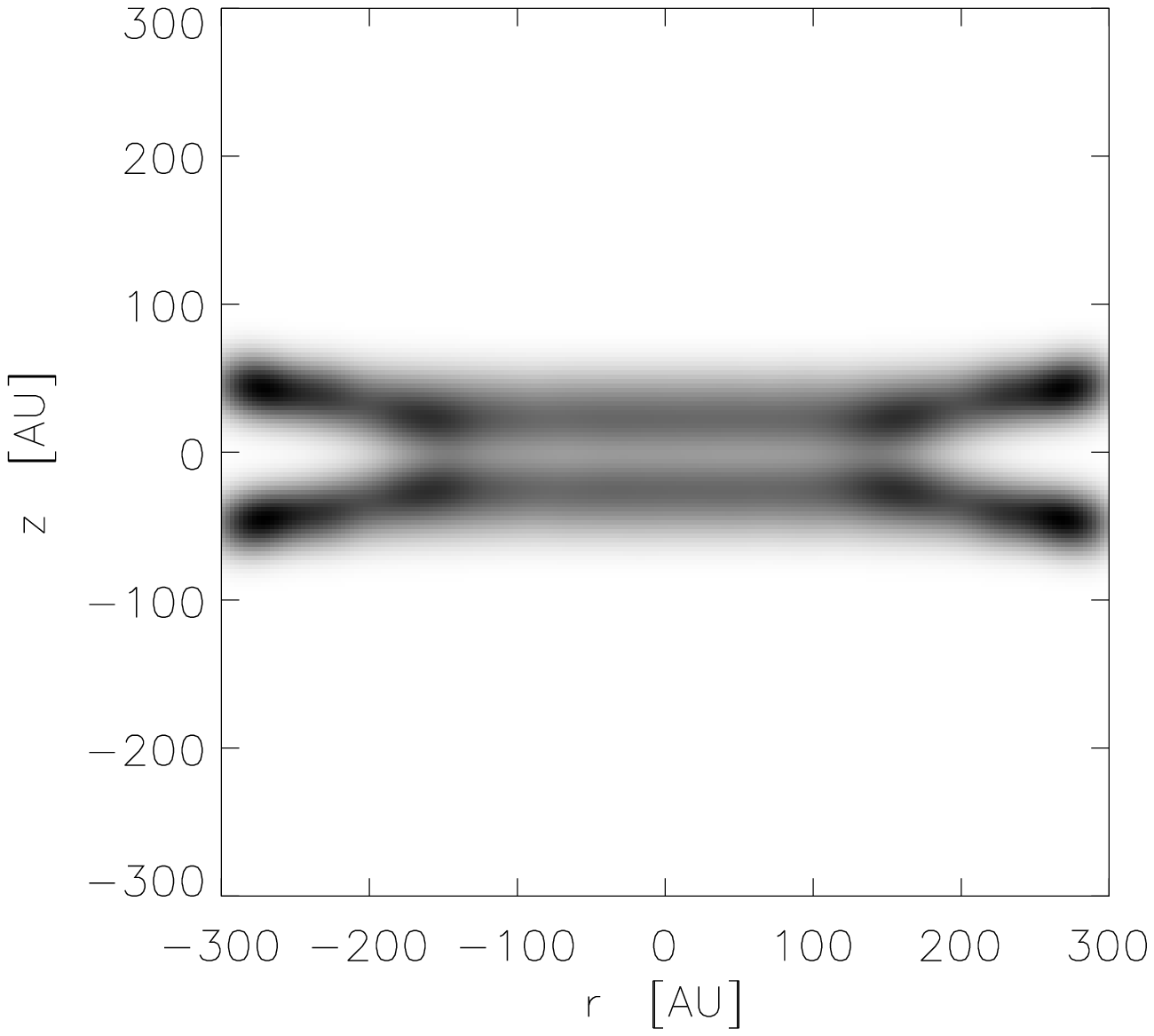}
   } 
  \caption{
    Different morphologies of an edge-on disc in the N band with $r_{\rm inner}=0.1 \au$.
    \emph{Top left:} $m_{\rm dust} = 3.2\times10^{-6}$, $r_{\rm outer}=300\au$, $\tilde{h}=12$, $\epsilon=0.001$.
    \emph{Top right:} $m_{\rm dust} = 3.2\times10^{-4}$, $r_{\rm outer}=300\au$, $\tilde{h}=20$, $\epsilon=0.1$.
    \emph{Bottom left:} $m_{\rm dust} = 3.2\times10^{-4}$, $r_{\rm outer}=300\au$, $\tilde{h}=20$, $\epsilon=0.1$.
    \emph{Bottom right:} $m_{\rm dust} = 3.2\times10^{-4}$, $r_{\rm outer}=400\au$, $\tilde{h}=8$, $\epsilon=1.0$. 
    All images feature a linear colour scale.
  }
  \label{img_N_zoo}
\end{figure}
In our chosen models, the re-emission in these bands dominates the scattered stellar flux by several orders of magnitude.
Morphologically, it is the transition region from the `dark lane' type of brightness distribution as seen in the near-infrared (see Section \ref{sect_nir}) to the extended and elongated typical for (sub)millimetre images.
In our edge-on images the typical central dark lane is very thin, in the order of one resolution element of the Giant Magellan Telescope.
The major difference to millimetre images of edge-on discs is that the flux due to re-emission is concentrated in the central region while at increasing wavelengths the emission becomes increasingly extended as cooler material dominates the emission. 

Based on our model we predict various morphologies that can be observed in this wavelength regime. 
Possible scenarios from within our parameter range include a single central peak, two elongated emission regions separated by a dark lane, three emission maxima in the midplane, a four-legged x-like shape and others more.
Fig.~\ref{img_N_zoo} gives an impression of the variety of possibilities.
The reason for this zoo of different morphologies resides in the fine-tuning of the optical depth.
Depending on the total mass and its distribution, different regions of the disc are rendered opaque enough for a specific choice of values  parameters such that their radiation contributes to an image (or not).
In most cases $\tau_\lambda$ is close to unity within an order of magnitude.

\begin{figure}
  \resizebox{\hsize}{!}{
    \includegraphics[width=\textwidth,angle=270]{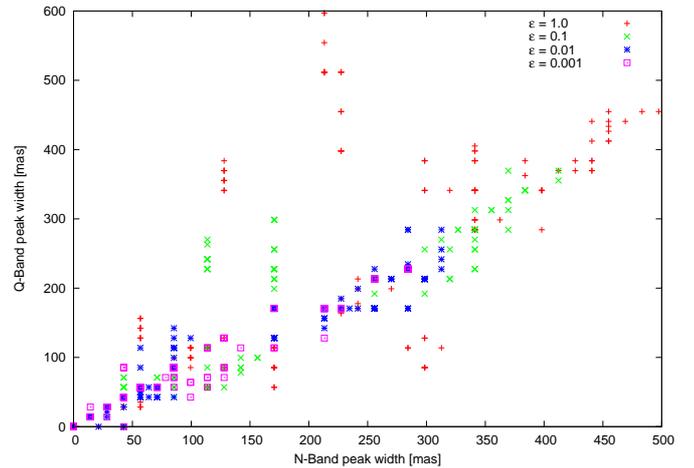}
   } 
  \caption{
    FWHM of the height of the emission in N band versus the same in Q band 
    for different values of the depletion parameter $\epsilon$.
    A general tendency of a linear relation between the FWHMs for smaller values of $\epsilon$ can be observed.
  }
  \label{plot_NQ_all}
\end{figure}

However, the overall width of the brightness distribution turns out to be a valuable tracer for dust grain growth and sedimentation in the edge-on case (see Fig. \ref{plot_NQ_all}).
A linear relation between the FWHM of the emissions in the two bands is observed.
For all possible combinations of parameters in our framework, the scattering of the Q band peak width vs. the N band peak width is strongest for a large depletion parameter and ceases for later evolutionary stages.

Introducing a straight line as a fit, the chi-square ($\tilde{\chi}^2$) is a function of the depletion parameter $\epsilon$ as shown in Table \ref{tab_epsvsstdev}.
\begin{table}
  \begin{tabular}{ccc}
    \hline
    Depletion $\epsilon$ & linear slope & $\tilde{\chi}^2$ \\
    \hline \hline
    1.0   & 1.0 & 10715 \\
    0.1   & 0.9 &  1820 \\
    0.01  & 0.8 &   381 \\
    0.001 & 0.8 &   191 \\
    \hline
    \end{tabular} 
    \caption{
      Linear slope and $\tilde{\chi}^2$ of the N/Q band width of the models.
    }
  \label{tab_epsvsstdev}
\end{table} 
The reason for this increasing linearity is the fine-tuning of the optical depth.
Re-emitted radiation received from the object originates at regions with an optical depth smaller than unity.
For large values of  $\epsilon$ the disc dust mass is distributed over a larger volume.
Thus, the iso-surface with $\tau_{\rm N band}$ has more morphological freedom.
Consequently, the width of the brightness distribution of one disc model picked at random depends more on the setting of the other parameters than this is the case for smaller values of $\epsilon$.
There the bulk of matter is concentrated close to the midplane and the shape of the regions seen is less subject to a specific choice of parameters:
The density distribution of the large grain population determines the width of the brightness distribution alone, not the interplay between two grain populations.
The specific value of the slope  in Table \ref{tab_epsvsstdev} is roughly determined by the wavelength dependency of $\tau$ and the vertical density increase of the large dust grain population as given by Eqn. \ref{eqdiscden}.
Increasing the observing wavelength from $10\mum$ to $20 \mum$ allows a deeper look into regions of higher disc densities.
For $\epsilon=0.001$ a slope of 0.8 implies that the Q band image is a little thinner than the N band image.
This means that the spatial gain by reduced optical depth is almost counterbalanced by the increased density in deeper layers.
 
Within our parameter space the FWHM-relation of the emission thickness in N and Q band for one disc depends  on the total dust mass.
An analysis of disc models shows, that they group in Fig. \ref{plot_NQ_all} according to their total dust mass, not their relative scale height. 
This is also true for later evolutionary stages.

Consequently, the evolutionary stage of a single observed disc can be inferred by comparing the extent of the emission in the N and Q band only.
The precise value of this relation will depend on the global geometry of the disc.
Nevertheless, a peak width ratio of $\sim 0.8$ from Q to N band is a good indicator for discs with most of their masses settled in large dust grains close to the midplane.

\subsection{Q and N band face-on: Turbulence digs in} 
\label{sec_NQ_gap}

\begin{figure}
  \resizebox{\hsize}{!}{
    \includegraphics[width=0.5\textwidth]{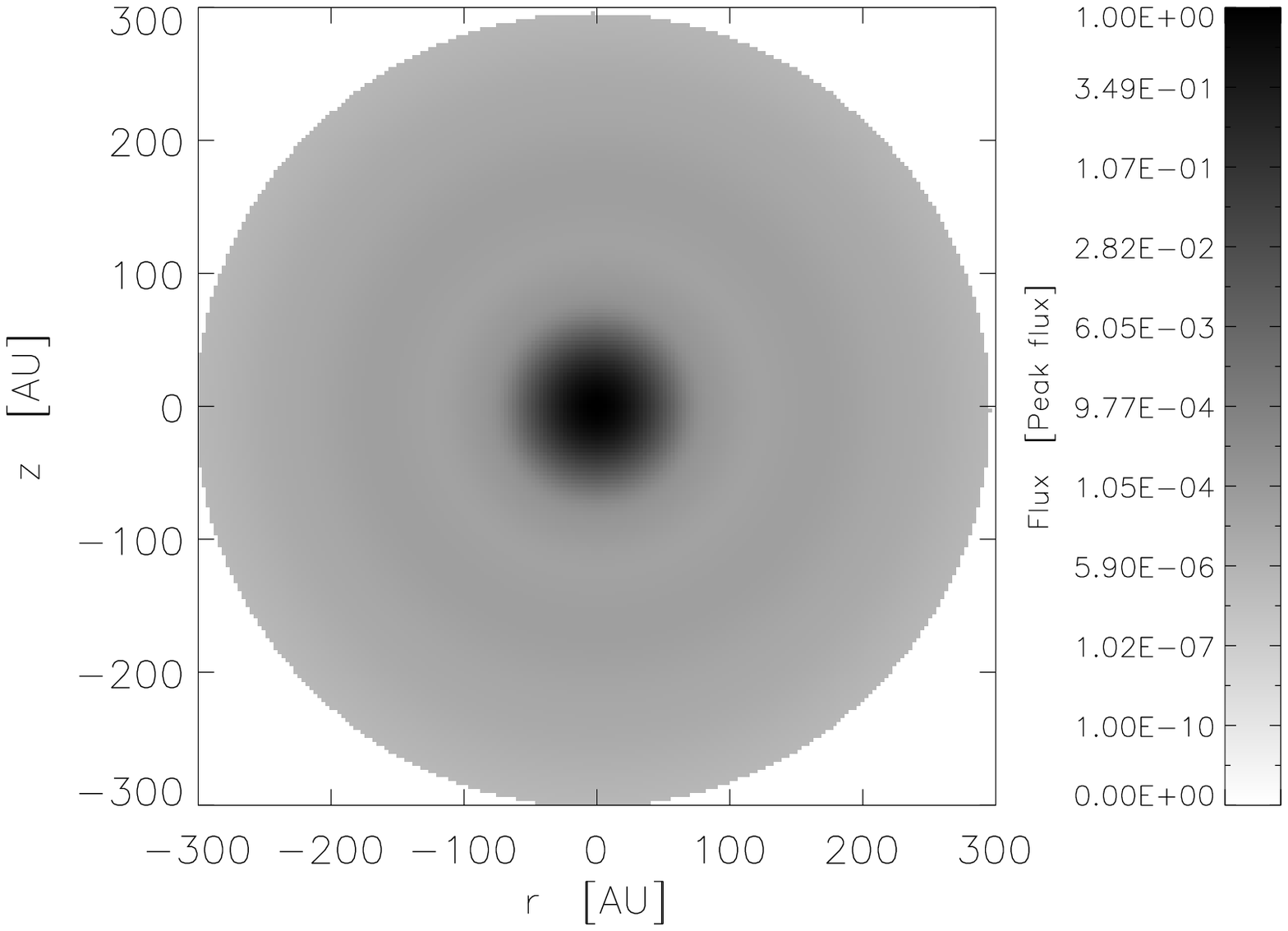}
  }  
  \resizebox{\hsize}{!}{    
    \includegraphics[width=0.5\textwidth]{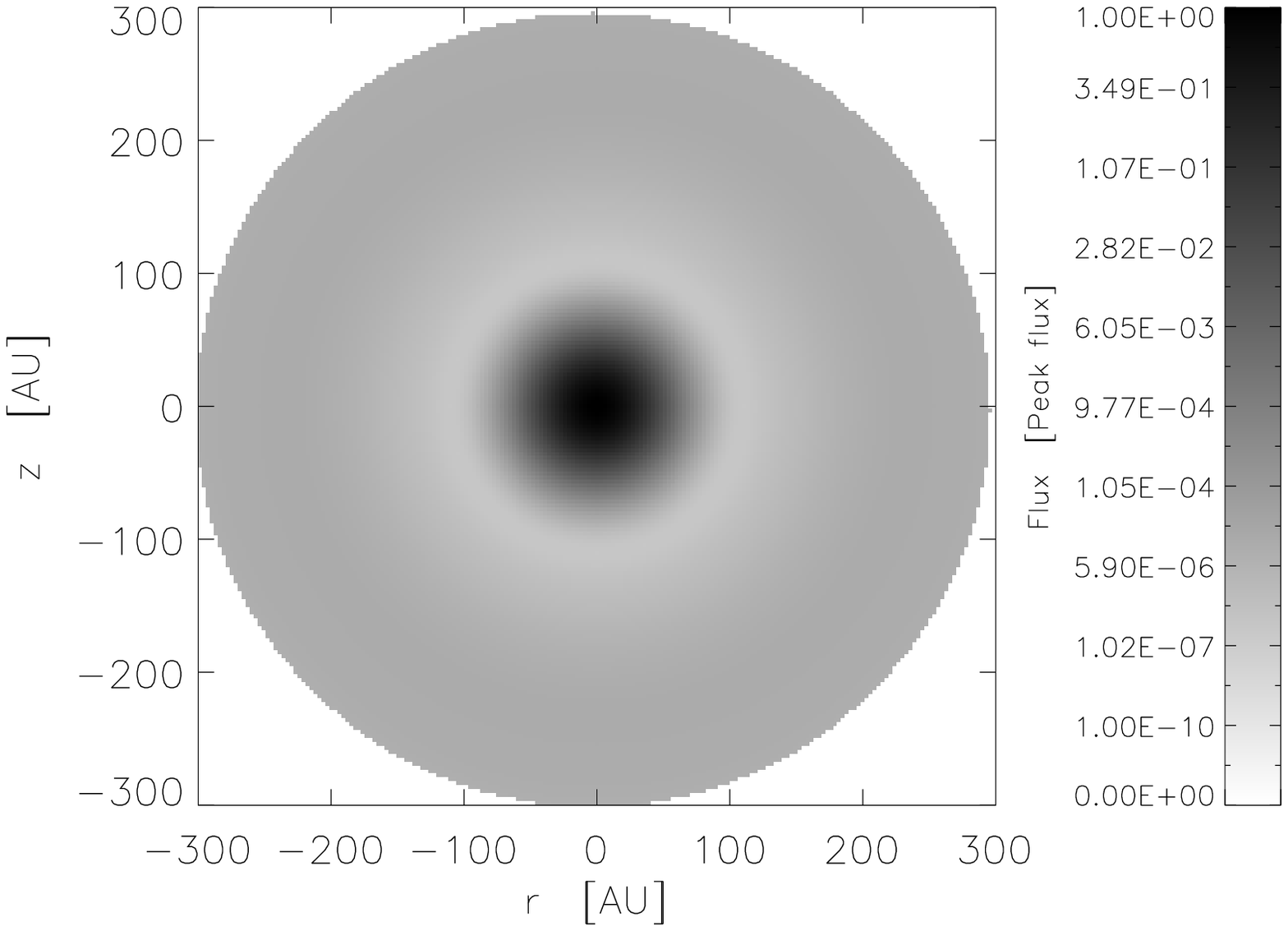}
  } 
  \resizebox{\hsize}{!}{
    \includegraphics[width=\textwidth]{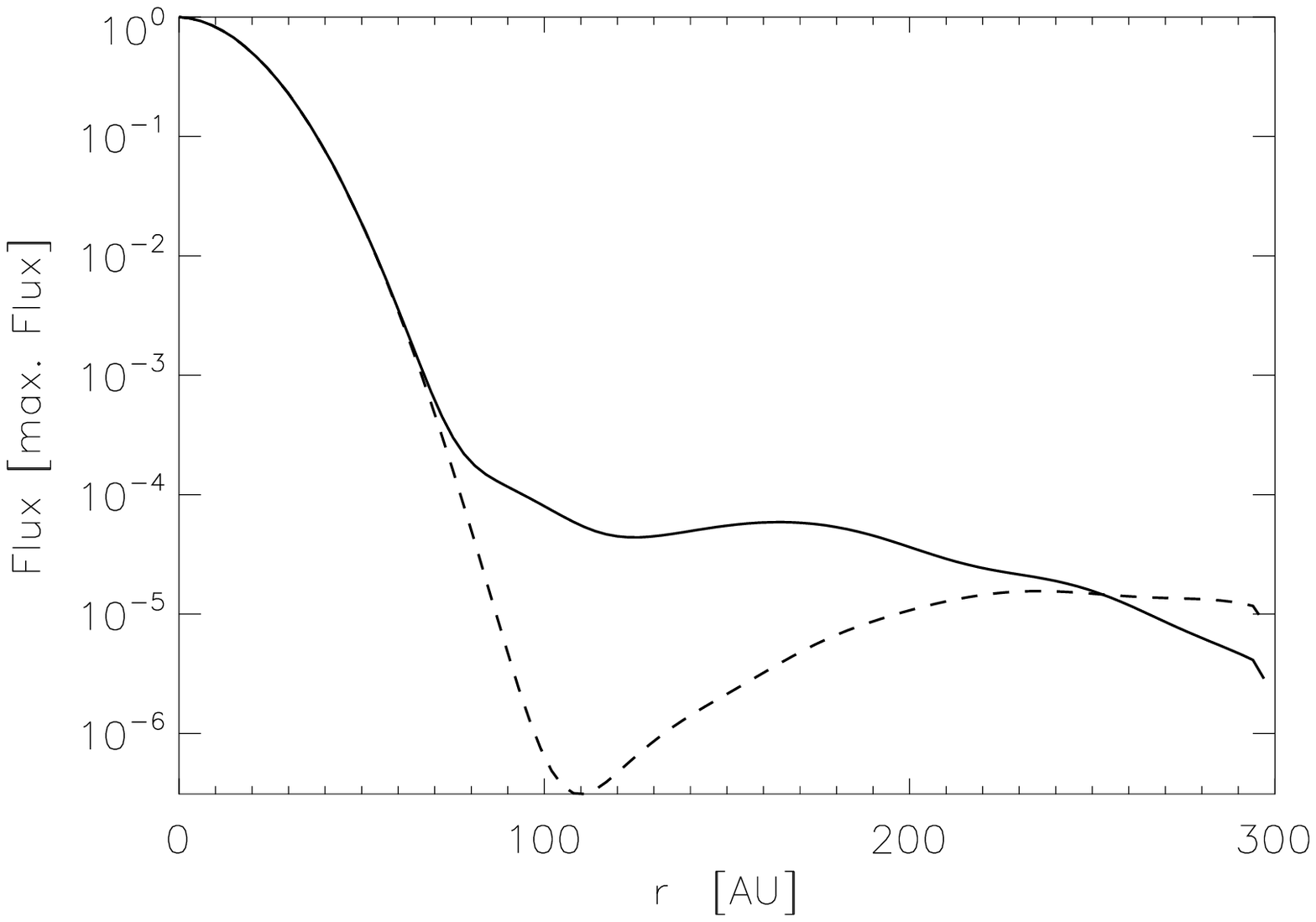}
  } 
  \caption{
    Face-on images in the N band and radial profiles of the brightness distribution.
    \emph{Top:} Fiducial model with a scale height relation of $\tilde{h}=8$.
    \emph{Centre:} Fiducial model with a scale height relation of $\tilde{h}=20$.
    \emph{Bottom:} Radial profiles.
      The solid lines corresponds to the first image, the dashed line to the second.
    All images have been convolved with a PSF of 30 AU.
    The colour scale is chosen as $\sim S_\nu^{0.1}$. 
  }
  \label{img_FO-N}
\end{figure}
This Section is dedicated to investigate the possibility to constrain the relative scale height $\tilde{h}$.
Fig.~\ref{img_N_zoo} already indicates that images in the N and Q band are sensitive to the fine-tuning of the density distribution.
Fig.~\ref{img_FO-N} illustrates this in the N band for the case of face-on discs.
In this regime the observed flux originates predominantly from re-emission of hot dust at $\sim 300\K$.
The marker for the relative scale height is the gap in the flux distribution at intermediate disc radii of $\sim100\au$.
The strength of this emission gap depends directly on the relative scale height $\tilde{h}$.
For large $\tilde{h}$, i.~e. a small scale height of the large dust grain population, the gap is clearly visible.
It ceases for smaller values of $\tilde{h}$.
The effect also depends on the observing wavelength.

\begin{figure}
  \resizebox{\hsize}{!}{
    \includegraphics[width=\textwidth]{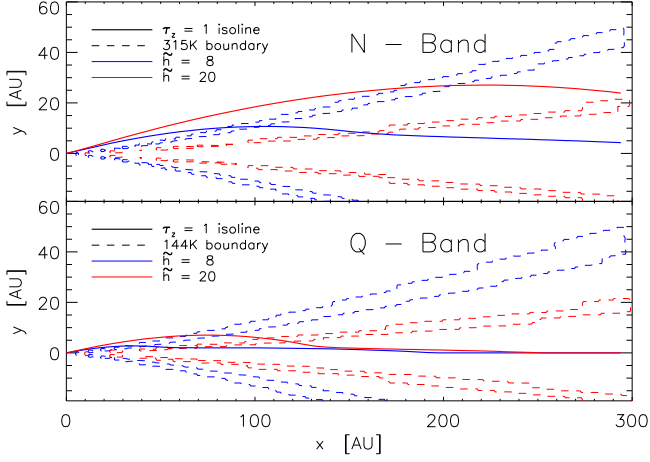}
  } 
  \caption{ 
  	Temperature distribution and isolines of constant optical depth $\tau=1$ in the N band \emph{(top)} and Q band \emph{(bottom)} for the fiducial model.
	Different colours refer to different relative scale heights $\tilde{h}$.
	Note the inclusion/exclusion of the 315K - emission region in the N band of the large-grain population
  }
  \label{plot_tau_temp}
\end{figure}
The nature of this apparent gap can be understood by analysing the spatial temperature distribution together with the optical depth.  
Fig.~\ref{plot_tau_temp} shows temperature contour lines of the large-grain population enclosing regions with temperatures higher than the black-body emission peak in the Q and N band.
This emission peak occurs for the Q band at 144\,K and in the N band at 315\,K.
Regions outside are cooler. 
Closer to the midplane the reason for this is the high optical depth such that the material there cannot be heated effectively by stellar radiation.
Far above the midplane the density of the disc is too low to assign a meaningful temperature (cf. Fig. \ref{plot_dens}).
Fig.~\ref{plot_tau_temp} also shows the $\tau_{\rm Q,N}=1$ isoline for the two bands obtained by integrating perpendicular to the midplane, starting at $z=\infty$.
For both bands and the two extreme values for $\tilde{h}$, the high temperature regions of the small grain population are far above the $\tau_{\rm Q,N}=1$ line and are not shown in the plot.
In both bands, these regions most efficiently contribute to the face-on image of the disc for any value of $\tilde{h}$.
These regions are the origin of the flux peak in the centre, cf. Fig. \ref{img_FO-N}.
 
In the Q band also the large-grain population contributes  to the image for both values of $\tilde{h}$.
Unconvolved face-on images show that the surface brightness of the small-grain population is about an order of magnitude larger.
The emission region of the small grain population is close to the rotation axis of the system.
Seen from above, all emission of the small grain population is seen piling up in a narrow region while the emission regions of the large grain population are stretched above the midplane.
In the edge-on case, this geometrical effect yields the opposite result:
The emission from the large grain population is seen in a smaller area and thus contributes more to the image than the small grain population.
  
Matters are different in the N band.
An optical depth of $\tau_{\rm N}=1$ is reached at higher altitudes from the midplane than in the Q band.
For $\tilde{h}=20$ and small radii, the emitting region is below the $\tau_{\rm Q}=1$ isoline.
This leads to a more dimmed emission in the N band for middle radii which results  in the  brightness gap at these radii.
The isolines are just indicating how deep face-on observations penetrate into the disc.
Wien's law only allows to determine those grains and disc layers which emit most efficiently in a given wavelength band \citep[e.g.][]{2006A&A...456..535S}.
Many more grains with lower or higher temperature can also provide the dominating flux at the wavelengths in question.
However, in Fig.~\ref{plot_tau_temp} all regions with higher temperatures are included by the temperature contour lines.
Considering that the emitted energy is proportional to $\sim\,T^4$, the reasoning in the previous paragraph remains robust.
  
\begin{figure}
  \resizebox{\hsize}{!}{
    \includegraphics*[width=\textwidth,angle=270]{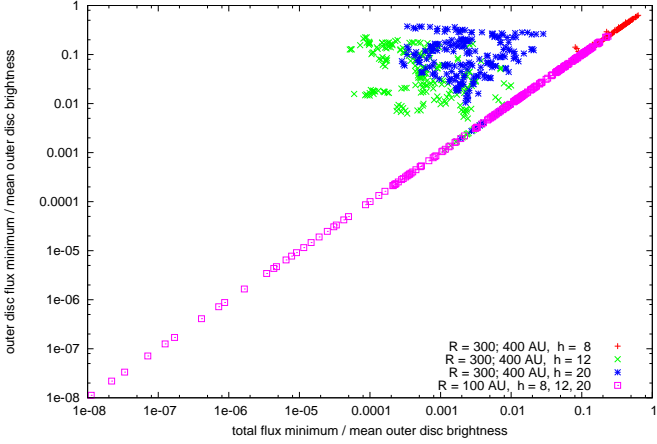}
   } 
  \caption{
    Classification of the apparent gap in the brightness distribution of edge-on seen in the N band.% $\lambda=10\mum$.
    Shown is the total brightness minimum in units of the average disc brightness in the outer $\slantfrac{1}{3}$ of the disc versus the minimum brightness in the outer $\slantfrac{1}{3}$ of the disc.
    Colour coded are model sets with different parameter values.
  }
  \label{plot_gapclass}
\end{figure}
\begin{figure} 
  \resizebox{\hsize}{!}{
    \includegraphics*[width=\textwidth,angle=270]{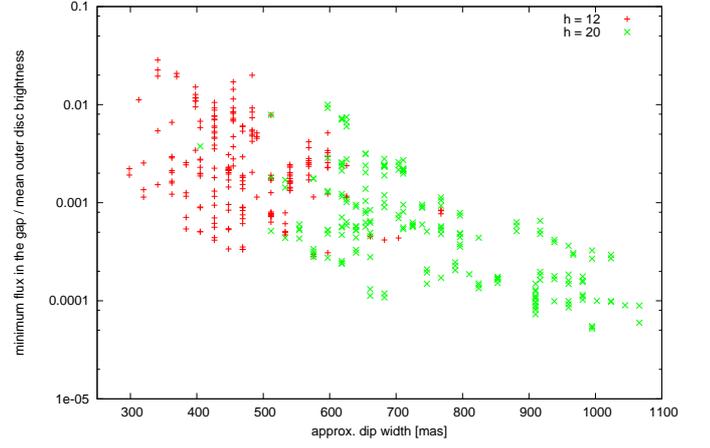}
   } 
  \caption{   
    Classification of the brightness gap of face-on seen discs at $\lambda=10\mum$.
    Plotted on a log-linear scale is the total brightness minimum in units of the average disc brightness in the outer $\slantfrac{1}{3}$ of the disc versus the gap width.
    Colour coded are models with different scale height relation $\tilde{h}$.
    Models featuring $\tilde{h} = 8$ do not exhibit a gap, see Fig. \ref{plot_gapclass}.
  }
  \label{plot_gapwidth}
\end{figure}

Fig. \ref{plot_gapclass} shows that the contrast of the brightness gap is a function of the relative scale height for two thirds of our parameter space.
Plotted is the total brightness minimum in units of the average disc brightness in the outer third of the disc versus the minimum brightness in the outer third of the disc also in units of the average outer disc brightness.
Models featuring a gap have a flux minimum within the first two thirds of the disc and are located left of the line with unity slope.
Models without an inner gap appear on the line with unity slope.
The gap is typically located at radii $r \approx 100\au$.
Discs with an outer disc radius of $r_{\rm outer}=100\au$ consequently do not exhibit a gap.
In Fig. \ref{plot_gapclass}, these discs are all located on the line with unity slope.
We define the width of the gap as the radial distance between the location of the first brightness maximum after the gap and the location where the brightness distribution close to the centre reaches the maximum value.
Fig. \ref{plot_gapwidth} shows the minimum flux in the gap as a function of this gap width.
Models with very pronounced gaps are thus located in the lower right part of the plot.
Shown are models with different values of $\tilde{h}$.
The strength of the gap does not correlate  with $\tilde{h}$ as can be seen from Fig. \ref{plot_gapwidth}.
Yet, the width of the gap does.
Our analysis further shows that the gap width does not correlate with the outer disc radius $r_{\rm outer}$.

\section{Other markers: Discussion}
\label{sec_discussion}
Our study also shows, that  markers others than those presented in Section \ref{sec_results} are not sufficient to constrain grain growth and coagulation.
However, we discuss them in this section to shed further light on the nature of the images of circumstellar discs.

\subsection{Radial profiles in the millimetre regime at low inclinations}
\label{mm_radprof_faceon}

The observed flux in (sub-)millimetre images stems from dust grain re-emission at these long wavelengths.
The optical depth  is small rendering all dust particles visible. 
As our models are rotationally symmetric, it suffices to consider the radial brightness profile instead of the whole image.
However, those profiles do not provide means by which discs at different stages of their evolution could be distinguished. 
While the absolute value of the radial profile of the surface brightness distribution depends on the evolutionary stage of the disc, the shape of the profile does not.
This renders the radial profile for $\theta<90^\circ$ degenerate with respect to $m_{\rm dust}$ and $\epsilon$.
This behaviour agrees with the findings of Section \ref{sec_millimeter}. 

\subsection{Dust lane chromaticity in the near-infrared}
\label{sect_nir}
In the near-infrared, the typical appearance of a circumstellar disc seen edge-on is a bipolar structure that is intersected by a dark lane.
The width of the dark lane is an effect of optical depth and wavelength dependent. 
In addition, at wavelengths between $2\mum$ and $10\mum$, radiation due to thermal re-emission of the dust grains becomes increasingly important in our disc models and is in the mid-infrared by far the dominant contributor to the total flux.
As the optical depth is a function of the disc mass, one expects the width of the dark lane to be dependent also on the depletion of the upper disc layers.
As grains grow and rain down towards the midplane, the upper layers become less opaque and
might affect the width of the dark lane as an indicator for $\epsilon$.
Increasing the observing wavelength unveils deeper layers in the disc and the large dust gain population (cf. Fig. \ref{plot_dens}) increasingly influences the appearance of the disk.

\begin{figure}
  \resizebox{\hsize}{!}{
    \includegraphics[width=\textwidth,angle=270]{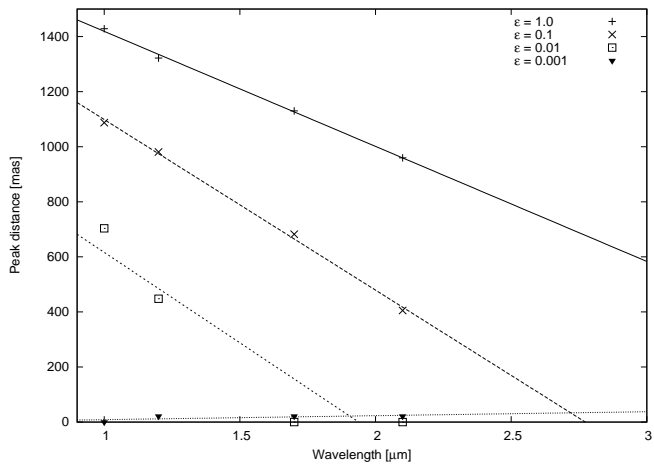}
   } 
  \caption{
    The chromaticity and dependency of the width of the dark lane on the depletion of upper disk layers in the NIR wavelength regime.
    For longer wavelengths and smaller $\epsilon$ the width is generally smaller.
    Shown also are straight line fits.
  } 
  \label{chromgood} 
\end{figure} 
Fig.~\ref{chromgood} shows the width of the dark lane for the fiducial model. 
As expected, the width of the dark lane decreases with both, wavelength and with progressing depletion of the upper disc layers.
The chromaticity itself is a function of $\epsilon$ (otherwise the fitted lines would be parallel): 
For smaller $\epsilon$ the chromaticity of the dark lane is less steep.
\begin{figure}
  \resizebox{\hsize}{!}{
    \includegraphics[width=\textwidth,angle=270]{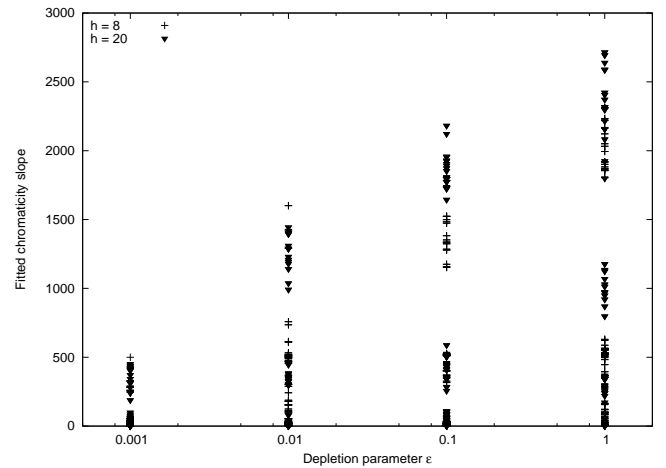}
   } 
  \caption{
    Scatter plot of the chromaticity-slope of the dark lane for different $\epsilon$ and different fit ranges.
  }
  \label{chrombad}
\end{figure}
 
Moreover, the chromaticity's dependence on $\epsilon$  provides means to identify dust grain growth and settling uniquely via near- to mid-infrared imaging only.
In Fig. \ref{chrombad} all chromaticity slopes $m_\epsilon$ for all possible parameter combinations within our framework are given.
For any given $\epsilon$, the slope obtained is either zero or scatters within a certain range.
The depletion of the upper disc layers could be inferred from the slope if a unique relation between $\epsilon$ and the chromaticity slope could be observed.
This is not the case.
The same behaviour is observed when not only images in the I, J, H, and K bands but also in the L and M bands are considered.
These two bands mark the transition region between images dominated by scattered stellar light and dust grain reemission.
This is independent of the resolution used to image a given circumstellar disc in the NIR. 

\subsection{Near-infrared and millimetre}
\begin{figure}
  \resizebox{\hsize}{!}{
    \includegraphics[width=\textwidth,angle=270]{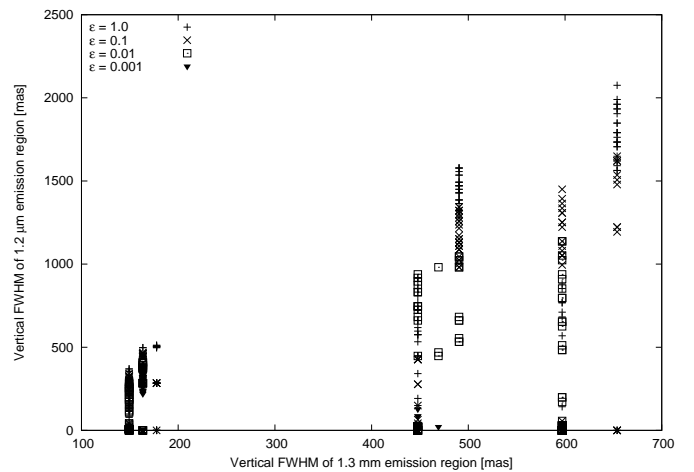}
   } 
  \caption{
    Width of the dark lane as function of the FWHM of the millimetre emission.
    Colour coded is the depletion parameter $\epsilon$.
    A clear dependency between $\epsilon$ and the width of both is not seen.
  }
  \label{mmvsNIR}
\end{figure}
One might expect that a combination of millimetre images and near-infrared images yields tracers for dust grain growth and sedimentation just as the combination of images in the N and Q.
Fig.~\ref{mmvsNIR} gives the relation between the FWHM of the vertical extend of the spatial brightness distribution in the millimetre regime and the width of the dark lane in the J Band.
The high dust densities in the disc mid-plane are responsible for  the dark lane in the NIR and the elongated spatial brightness distribution in the millimetre regime.
Thus, a relation between the width of the dark lane and the millimetre emission region could hint to the disc's evolution.
However, as Fig. \ref{mmvsNIR}  shows, no correlation between the two quantities is observable.
 
\subsection{Degeneracy between parameters and panchromatic studies}
We point out that if all disc parameters in our parameter-space are known, in most cases the chromaticity of the dark lane uniquely determines the evolutionary state, as illustrated by the fiducial model (cf. Section \ref{sec_fidu}).
In general, it is not possible, though, to determine the evolutionary stage of a disc when other parameter values remain unconstrained as well.

As a simple counter-example, consider a system that is optically thin at all near-infrared wavelengths, and thus a dark lane is not visible and no chromaticity can be seen.
In our framework, his happens for discs with either a low total dust mass, a low scale height relation $\tilde{h}$, a high depletion parameter $\epsilon$, or combinations thereof.
The total dust mass can be constrained by the systems' total flux in the millimetre regime where the disc is generally optically thin.
However, to constrain $\tilde{h}$ a certain total dust mass is required.
Otherwise the effects of $\tilde{h}$ are rendered indistinguishable.

\subsection{The general face-on case}
\begin{figure} 
  \begin{center}
  \resizebox{\hsize}{!}{
    \includegraphics[width=0.5\textwidth,angle=270]{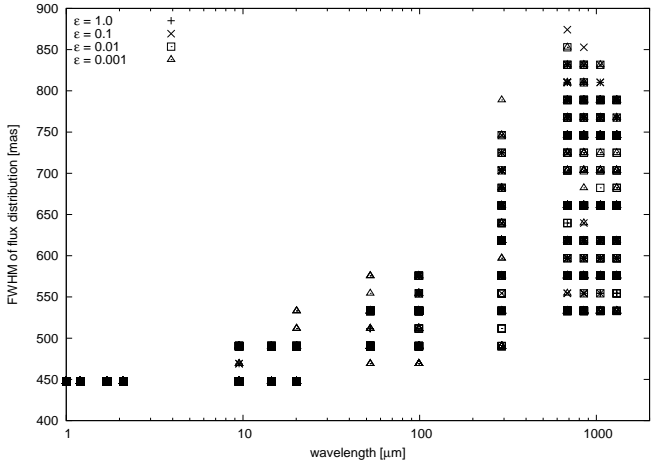}
   } 
  \caption{
    The FWHM of the spatial intensity distribution of disc observed face-on as function of the wavelength.
    The depletion parameter $\epsilon$ is encoded with different symbols.
  } 
  \label{faceonplots}
  \end{center}
\end{figure}
As shown for millimetre images in Section \ref{sec_millimeter}, the ability to constrain dust sedimentation by single imaging is lost for $\theta<90^\circ$.
For images in the near-infrared and mid-infrared the situation is similar.
In Fig.~\ref{faceonplots} plots of the FWHM of the disc's brightness distribution seen edge-on at different wavelengths are given.
At short wavelengths the emission is very small as in the NIR the star is the dominant contributor to the image flux.
Images exhibit an apparently broader flux distribution at larger wavelengths since the origin of the observed flux shifts from scattering to re-emission.
The size of the re-emission regions growths as the observing wavelengths gets larger and cooler parts become more important contributors.
Whereas in Fig.~\ref{plot_NQ_all} a grouping of data points with smaller $\epsilon$ to a linear relation between the N and Q band emission width can be observed no such relation is seen in a NIR/millimetre comparison.
Even if a parameter is well constrained by some observation no specific behaviour of data points with the same depletion parameter $\epsilon$ can be seen.
 
This observation does not contradict the findings of \cite{2004A&A...421.1075D}.
In their work, dust grain growth and sedimentation have not been modelled with a parametrised model but including radius dependent sedimentation.
In our model framework we do not allow for radial depended dust settling.
Quicker dust grain growth at smaller radii renders the upper disc layers in the inner disc more transparent and allows to heat the outer disc more efficient. 
The radial brightness profile would be shifted and/or spread outward.
In contrast, the upper layers deplete all at the same rate in our framework.
Hence the radial brightness profile is unaltered.
The latter is only true for the relative profile, that is a profile normalised to some other disc inherent quantity.
For most practical reasons this would be the total disc flux.
The absolute value of fluxes changes with a the evolution of the disc (cf. Section \ref{mm_radprof_faceon}).
  
\section{Conclusions}
In this study we explore the effects of dust grain growth and sedimentation on images of circumstellar discs at various wavelengths and inclinations.
As the prediction of observable quantities of circumstellar discs is highly non-linear and harbours several degenerate parameters, we employ a simplified, parametrised model.
This allows us to discuss the general effects of the depletion of upper disc layers by grain growth and dust sedimentation using radiative transfer simulations in the near-infrared, the mid-infrared as well as in the (sub-)millimetre regime.
In the advent of telescopes providing high spatial resolution we focus on images at these wavelengths thus extending the existing studies on SEDs.
Although our approach is based on certain simplifications, we identify tracers that uniquely allow to constrain the evolutionary stage of a disc.

In particular, our findings have direct implications on future observations:
\begin{itemize}
	\item The width of the dark lane in NIR images of edge-on discs is degenerate with respect the degree of depletion of upper disc layers from large grains and observing wavelength.
	Although this wavelength and the degree of depletion do not exactly mirror each others impact on the width of the dark lane, a given width lane can be reproduced adjusting both parameters (see Section \ref{sec_fidu}).
	\item High spatial resolution also needs to be accompanied by high sensitivity in the case of millimetre images.
	 Only ALMA will provide the required resolution and sensitivity to identify grain sedimentation in millimetre images of edge-on seen discs.
	Settled disc show a relative surface brightness decrease of $\sim10^{-4}$ from the mid-plane to 150 AU above the disc while non-settled discs show a decrease of only $\sim10^{-2}$ (see Section \ref{sec_millimeter}).
	\item If the disc is not seen edge-on, then this distinction between settled discs and non-settled discs cannot achieved by millimetre observations, if the settling is independent of the radial position in the disc.
	 \item 
    A comparison of the FWHM of the brightness distribution in N and Q band images allows to constrain the degree of dust sedimentation in a circumstellar disc if seen edge-on.
    Evolved discs in our parameter space show a ratio of the Q band FWHM to the N band FWHM of $\sim0.8$ which depends only very weakly on other disc properties.
    Young discs however show a ratio that depends strongly on other disc parameters and scatters for various combinations of parameter values around the average FWHM ratio of $\sim 1.0$ (see Section \ref{sec_NQ_faceon}).
    \item The relative sedimentation height $\tilde{h}$ to which particles settle can be determined using face-on observations in the N and Q band. 
    	Due to the complex interaction between the spatial structure of the optical depth and the temperature distribution discs with settling of large grains close to the mid plane exhibit a gap in the observed radial brightness distribution at typically $\sim 100\au$.
	However, the effect is only seen in discs considerably larger than $\sim100\au$ (see Section \ref{sec_NQ_gap}).
\end{itemize}  

Furthermore, we discuss ambiguities that prevent to achieve the same results for images in other wavelength regimes, namely the near-infrared.
As a next step  a disc parametrisation should be chosen that allows to incorporate radially dependent sedimentation.
Also, modelling of images of an individual object will provide a better insight in its particular evolutionary stage.
The knowledge gained the framework of the present paper forms a non-comprehensive basis for further analysis of dust grain growth and sedimentation effects on images.

\begin{acknowledgements}
J. S. thanks C. Dullemond for enlightening discussions.
J. S. acknowledges support by the DFG through the research group 759 ``The Formation of Planets: The Critical First Growth Phase''.
\end{acknowledgements} 
 
\bibliographystyle{aa}
\bibliography{sett_coag.bib}

\end{document}